# Specifics of formation tax revenues and ways to improve it in Georgia

George Abuselidze and Rusudan Zoidze

Batumi Shota Rustaveli State University

Abstract

In the research there is reviewed the peculiarities of the formation of tax revenues of the state budget, analysis of the recent past and present periods of tax system in Georgia, there is reviewed the influence of existing factors on the revenues, as well as the role and the place of direct and indirect taxes in the state budget revenues. In addition, the measures of stimulating action on formation of tax revenues and their impact on the state budget revenues are established.

At the final stage, there are examples of foreign developed countries, where the tax system is perfectly developed, where various stimulating measures are successfully stimulating and consequently it promotes mobilization of the amount of money required in the state budget.

The exchange of foreign experience is very important for Georgia, the existing tax model that is based on foreign experience is greatly successful.

For the formation of tax policy, it is necessary to take into consideration all the factors affecting on it, a complex analysis of the tax system and the steps that will be really useful and perspective for our country.

# საგადასახადო შემოსავლების ფორმირების თავისებურებები და მისი სრულყოფის გზები საქართველოში

გიორგი აბუსელიძე და რუსუდან ზოიძე

ბათუმის შოთა რუსთაველის სახელმწიფო უნივერსიტეტი

**ანოტაცია**

ნაშრომში განხილულია სახელმწიფო ბიუჯეტის საგადასახადო შემოსავლების ფორმირების თავისებურებები, საქართველოში არსებული საგადასახადო სისტემის ახლო წარსულისა და მიმდინარე პერიოდის ანალიზი. გამოკვეთილია საგადასახადო შემოსავლების ფორმირებაზე მოქმედი ფაქტორების გავლენა. ასევე პირდაპირი და არაპირდაპირი გადასახადების როლი და ადგილი სახელმწიფო ბიუჯეტის შემოსავლებში. გარდა ამისა, ჩამოყალიბებულია საგადასახადო შემოსავლების ფორმირებაზე მოქმედი სტიმულირების ღონისძიებები და მათი გავლენა სახელმწიფო ბიუჯეტის შემოსულობებზე.

დასკვნით ეტაპზე კი წარმოდგენილია უცხოეთის იმ განვითარებული ქვეყნების მაგალითები, სადაც სრულყოფილად არის განვითარებული საგადასახადო სისტემა, წარმატებით ფუნქციონებს სხვადასხვა მასტიმულირებელი ღონისძიებები და შესაბამისად, იგი ხელს უწყობს სახელმწიფო ბიუჯეტში საჭირო ოდენობის ფულადი სახსრების მობილიზებას. უცხოური გამოცდილების გაზიარება საქართველოსთვის მეტად მნიშვნელოვანია, აქამდე არსებული საგადასახადო დაბეგვრის მოდელები, რომლებიც უცხოურ გამოცდილებას ეფუძნება დიდი წარმატებით ფუნქციონირებს. საგადასახადო პოლიტიკის სრულყოფილად ფორმირებისათვის აუცილებელია გათვალისწინებული იქნეს მასზე მოქმედი ყველა ფაქტორის გავლენა, მოხდეს საგადასახადო სისტემის კომპლექსური ანალიზი და გადაიდგას ისეთი ნაბიჯები, რომელიც რეალურად სასარგებლო და პერსპექტიული იქნება ჩვენი ქვეყნისათვის.

**შესავალი**

საგადასახადო სისტემა ქვეყნის ეკონომიკური სისტემის მნიშვნელოვანი ელემენტია. სწორედ ამასთან დაკავშირებით საგადასახადო დაბეგვრის, როგორც ეკონომიკაზე სახელმწიფოს ზემოქმედების ინსტრუმენტის, სრულყოფისა და განვითარების პრობლემა არის ერთ-ერთი აქტუალური თემა ეკონომიკურ მეცნიერებასა და პრაქტიკაში. ამდენად აღნიშნული მოითხოვს შესწავლილი და გაანალიზებული იქნას სამამულო და საზღვარგარეთის ქვეყნების გამოცდილება, მეცნიერ-ეკონომისტთა და პრაქტიკოსთა შეხედულებები, თეორიები და წინადადებები.

ნაშრომის მიზანს წარმოადგენს საგადასახადო შემოსავლების ფორმირების თავისებურებების განსაზღვრა, საგადასახადო სისტემის ფუნქციონირებაზე მოქმედი ფაქტორების გამოვლენა და მისი გაუმჯობესების მიზნით საზღვარგარეთული პრაქტიკის გამოყენება; ამოცანას კი იმ გზების და საშუალებების მიგნება, რომელთა გატარება სრულყოფს საქართველოს საგადასახადო სისტემას და აქედან გამომდინარე დადებითად იმოქმედებს სახელმწიფო ბიუჯეტის შემოსავლებზე.

კვლევის საგანი და ობიექტი. კვლევის საგანია საქართველოს საგადასახადო სისტემა, ხოლო ობიექტი სახელმწიფო ბიუჯეტის საგადასახადო შემოსავლები.

თეორიულ და მეთოდოლოგიურ საფუძვლად გამოყენებულია გადასახადებსა და საგადასახადო სისტემის საკითხებზე მიძღვნილი სამეცნიერო ნაშრომები, ოფიციალური დოკუმენტები, საკანონმდებლო აქტები და სტატისტიკური მასალები.

1. საქართველოს სახელმწიფო ბიუჯეტის საგადასახადო შემოსავლების ფორმირების თეორიულ-მეთოდოლოგიური საფუძვლები

1.1 *საგადასახადო პოლიტიკის ფორმირების თავისებურებები და მისი გავლენა ბიუჯეტის საგადასახადო შემოსავლებზე*

სახელმწიფოს მიერ ფორმულირებული საგადასახადო პოლიტიკა და საგადასახადო სისტემა მნიშვნელოვან გავლენას ახდენს ქვეყანაში ეკონომიკური სისტემის ეფექტურ ფუნქციონირებაზე. ქვეყნის წინაშე არსებული სოციალური პრობლემების გადაწყვეტა, წარმოების განვითარება და მდგრად ეკონომიკური განვითარების მიღწევა, მხოლოდ გამართული და სტაბილური სოციალურ-ეკონომიკური პირობების შესატყვისი საგადასახადო სისტემის ფორმირების პირობებშია შესაძლებელი. საგადასახადო სისტემის სრულყოფასთან დაკავშირებული საკითხები ყოველთვის არის და რჩება ქვეყნის ეკონომიკური პოლიტიკის უმნიშვნელოვანეს საკვანძო საკითხად.

როგორც ვიცით, სახელმწიფოს განვითარების ეკონომიკური საფუძველია ფინანსური რესურსები, რომლებიც ძირითადად გადასახადებითა და მათი სახელმწიფო ბიუჯეტში გადახდით ფორმირდება. გადასახადები სახელმწიფოს მხრიდან ეკონომიკაზე ზემოქმედების მნიშვნელოვანი იარაღია. ზოგადად, საგადასახადო სისტემა იმგვარად უნდა იყოს ფორმულირებული, რომ იგი ადვილად შეეწყოს ცვალებად ეკონომიკურ გარემოს და დროის კონკრეტულ მომენტში ქვეყნის ეკონომიკური მიზნებისა და ამოცანების გათვალისწინება შეეძლოს.

პოსტსაბჭოური სისტემის პერიოდში, როცა სისტემაში შემავალი ყველა სახელმწიფოსათვის წინასწარ იყო განსაზღვრული თუ რა, რამდენი, სად და როდის ეწარმოებინათ, სად გაეყიდათ და რა მოცულობის შემოსავალი მიეღოთ, ფაქტობრივად ჩვენს ქვეყანას არ გააჩნდა საგადასახადო სისტემა მისი კლასიკური გაგებით, რაც ასევე იმითაც იყო განპირობებული, რომ წარმოების საშუალებებიც და წარმოებული პროდუქციაც სახელმწიფო საკუთრებას შეადგენდა. შესაბამისად, ყველა რესურსი, ადმინისტრაციული მეთოდების გამოყენებით, სახელმწიფოს ნებისმიერ დროს, ნებისმიერი მოცულობით შეეძლო მოეზიდა ბიუჯეტში.

საბჭოთა კავშირის დაშლის შემდეგ ქვეყანაში წარმოიქმნა საკუთარი საგადასახადო სისტემის ჩამოყალიბების აუცილებლობა, რომელსაც როგორც სახელმწიფო ინტერესები, ასევე თავისუფალი ბაზრის მოთხოვნებიც უნდა დაეკმაყოფილებინა. საქართველოს თანამედროვე საგადასახადო სისტემა 1991-1992 წლების მიჯნაზე, ქვეყანაში მიმდინარე ეკონომიკური გარდაქმნებისა და საბაზრო ურთიერთობებზე გადასვლის პერიოდში ჩამოყალიბდა.

საბაზრო ეკონომიკის პირობებში სახელმწიფო ბიუჯეტის შემოსავლებისა და გასავლების ფორმირების კანონზომიერებანი და მეთოდოლოგია ძირეულად განსხვავებულია. რეალურად, საგადასახადო ურთიერთობების სამართლებრივი რეგულირების გამოცდილების უქონლობამ, ეკონომიკურმა და სოციალურმა კრიზისმა ქვეყანაში, კანონმდებლობის შემუშავებისათვის არსებულმა შემჭიდროვებულმა ვადებმა, დიდი გავლენა მოახდინა საგადასახადო სისტემის ჩამოყალიბებაზე.

საქართველოს საგადასახადო სისტემა ძირითადად საზღვარგარეთის ქვეყნების გამოცდილების ბაზაზე იქმნებოდა, თუმცა აღსანიშნავია ის გარემოებაც, რომ გათვალისწინებული უნდა იყოს ეროვნული სპეციფიკის თავისებურებები. თუმცა, მიუხედავად აღნიშნულისა, საქართველოს საგადასახადო სისტემა საერთო სტრუქტურითა და აგებულების პრინციპებით მსგავსია მსოფლიო ეკონომიკაში გავრცელებული საგადასახადო დაბეგვრის სისტემებისა.

საგადასახადო სისტემის შემუშავებისას პირველ რიგში აუცილებელია საგადასახადო შემოსავლების სახეების, მათი ეფექტურად ამოღებისა და გაზრდის შესაძლებლობათა განსაზღვრა. გარდა ამისა, უნდა განისაზღვროს მათი ზემოქმედება ფასებზე, სტიმულებზე თუ სხვა მაკროეკონომიკურ მაჩვენებლებზე. როგორც ვიცით, გადასახადები ცალკეულ ბაზრებსა თუ ეკონომიკურ სექტორებზე განსხვავებულად ზემოქმედებს და შესაბამისად „საგადასახადო-საბიუჯეტო პოლიტიკის წარმატებით წარმართვა საგადასახადო სისტემის ისეთი მახასიათებლების გათვალისწინებას მოითხოვს, როგორიცაა: სტაბილურობა ელასტიურობა, ეფექტურობა, სამართლიანობა, მართვის სიმარტივე, მოხერხებულობა, საგადასახადო ტვირთის გონივრული საზღვრების დადგენა.

„იდეალური" საგადასახადო სისტემა რამდენიმე გადასახადზე და ფართო საგადასახადო ბაზაზე უნდა იყოს დაფუძნებული, რაც საბოლოო ჯამში, გადასახადების გადახდას და გადასახადების დაგეგმილი მოცულობით ამოღებას აადვილებს" (ფუტკარაძე, 2012).

საერთაშორისო რეიტინგით "Doing bussines 2019" ბიზნესის დაწყების სიმარტივით საქართველო მე-6 ადგილს ისეთი ქვეყნების შემდგომ იკავებს როგორებიცაა: ახალი ზელანდია, სინგაპური, ჰონ-კონგი და კორეის რესპუბლიკა, რაც თავის მხრივ საგადასახადო სისტემის მოქნილობას უსვამს ხაზს. 2011 წლის 1 იანვრიდან ძირეული ცვლილებები შევიდა საგადასახადო კოდექსში, რაც გულისხმობდა ბიზნესის სტიმულირებას და გამარტივებული საგადასახადო რეგულირების შემოდებას, დაფუძნდა საგადასახადო ომბუდსმენისა და პირადი საგადასახადო აგენტის ინსტიტუტი, დაწესდა დამატებული ღირებულების გადასახადის დაბეგვრის ახალი რეჟიმი, რაც გადასახადის კვარტალურ დეკლარირებას გულისხმობს, და რაც ყველაზე მნიშვნელოვანი და ძირეული ცვლილებაა, დამატებული ღირებულების გადასახადის გადამხდელის მიერ ზედმეტად გადახდილი გადასახადის ავტომატურ რეჟიმში უკან დაბრუნება გახდა შესაძლებელი, აქამდე მოქმედი კანონქვემდებარე აქტები კი 5 კანონქვემდებარე აქტად ჩამოყალიბდა.

ზოგადად, "ქვეყანაში საგადასახადო სისტემა, საგადასახადო პოლიტიკა ისე უნდა მოეწყოს, გადასახადების სახეები და განაკვეთები, საგადასახადო კონტროლი, საგადასახადო დავების გადაჭყვეტის წესები და საკითხები, გადასახადების გადამხდელებსა და სახელმწიფოს შორის ურთიერთობები ისე უნდა გვარდებოდეს, რომ ამ ურთიერთობებმა არათუ ხელი არ შეუშალოს თავისუფალ მეწარმეობას და კონკურენციის განვითარებას, არამედ პირიქით, ხელი შუწყოს მას, ისე რომ ყველას ვინც მეწარმეობას და, საზოგადოდ, ეკონომიკურ საქმიანობას ეწევა, გამონაკლისის გარეშე, თანაბარ პირობებში მოუწიოს მუშაობა" (ფუტკარაძე, 2012).

როგორც წესი, ნებისმიერი ქვეყნის საგადასახადო სისტემას გააჩნია ორი ძირითადი მიზანი: პირველ რიგში მან უნდა უზრუნველყოს სახელმწიფო ბიუჯეტის შემოსავლების ძირითადი ნაწილის ფორმირება, რომელიც აუცილებელია სახელმწიფოს მიერ განსაზღვრული ხარჯების დასაფარავად, ხოლო მეორეს მხრივ, საგადასახადო სისტემა სტიმული უნდა იყოს ეკონომიკური აქტივობისათვის, ინვესტიციების მოზიდვისათვის, წარმოების გაფართოება-განვითარებისა და შესაბამისად, აღნიშნულის კვალობაზე, საგადასახადო ბაზის შემდგომი გაზრდისათვის.

საქართველოს სახელმწიფო ბიუჯეტის შემოსავლებში ერთ-ერთი მნიშვნელოვანია საგადასახადო შემოსავლები, რომელიც სახელმწიფო ბიუჯეტის ფორმირების ძირითადი წყაროა (ცხრილი №1).

ცხრილი 1: საგადასახადო შემოსავლების %-ლი წილი სახელმწიფო ბიუჯეტში 2014-2019 წლებში (ათასი ლარი)

|  | 2014 წლის ფაქტიური შესრულება | 2015 წლის ფაქტიური შესრულება | 2016 წლის ფაქტიური შესრულება | 2017 წლის ფაქტიური შესრულება | 2018 წლის ფაქტიური შესრულება | 2019 წლის გეგმიური მაჩვენებელი |
|---|---|---|---|---|---|---|
| შემოსულობები | 9,157,085.1 | 9,891,079.8 | 10,374,024.0 | 11,618,665.1 | 12,693,422.2 | 12,863,750.0 |
| გადასახადები | 6,846,964.2 | 7,549,608.9 | 7,986,750.3 | 8,991,307.5 | 9,695,962.2 | 9,645,000.0 |
| %-ული წილი | 74.77 | 76.33 | 76.99 | 77.39 | 76.39 | 74.98 |

წყარო: შედგენილია ავტორის მიერ საქართველოს ფინანსთა სამინისტროს მონაცემებზე დაყრდნობით ცხრილი 1-ზე წარმოდგენილი მონაცემები ნათელ სურათს გვაძლევს იმის შესახებ თუ რაოდენ მნიშვნელოვანია ქვეყნის ბიუჯეტის ფორმირებისათვის საგადასახადო სისტემის სრულყოფა. გამომდინარე იქედან, რომ საბიუჯეტო შემოსულობებში საგადასახადო შემოსავლების წილი არც თუ ისე სახარბიელოა აუცილებელია სწორი და ქმედითი ნაბიჯები გადაიდგას საგადასახადო პოლიტიკის შემუშავებისას, რომ საბიუჯეტო შემოსავლების ძირითადი წყარო იყოს გადასახადებით მობილიზებული სახსრები და რა იყოს იგი დამოკიდებული სხვა წყაროებზე, როგორიცაა: ვალდებულებების აღება და ფინანსური და არაფინანსური აქტივების ცვლილება რადგან მან ზიანი არც საბიუჯეტო შემოსავლებს და არც გადასახადების გადამხდელებს მიაყენოს, რომლებიც ეკონომიკაში ქმნიან მატერიალურ დოვლათს მთლიანი შიდა პროდუქტის სახით.

ქვემოთ მოცემულ დიაგრამა №1 კი ასახულია 2014-2019 წლებში ბიუჯეტში მობილიზებულ საგადასახადო შემოსავლებში თითოეული გადასახადის %-ული წილი.

**დიაგრამა 1:** 2014-2019 წლების სახელმწიფო ბიუჯეტის შემოსავლები გადასახადების მიხედვით (ათასი ლარი)

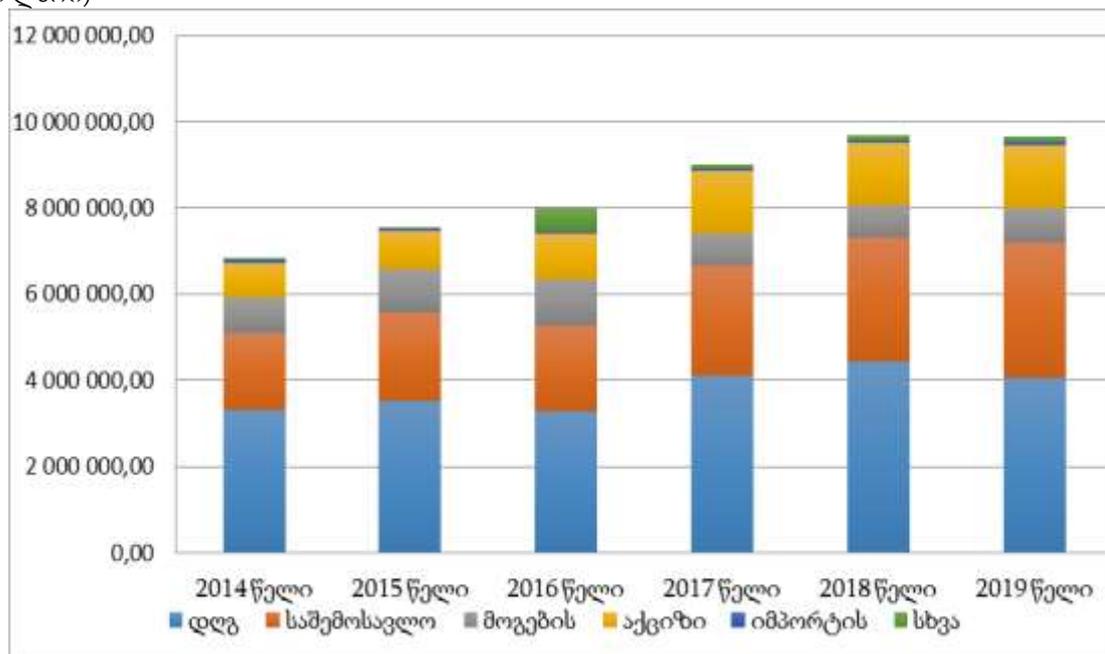

**წყარო:** შედგენილია ავტორის მიერ საქართველოს ფინანსთა სამინისტროს მონაცემებზე დაყრდნობით როგორც ვხედავთ საგადასახადო შემოსავლებში დიდი წილი დამატებული ღირებულების, საშემოსავლო და მოგების გადასახადებზე მოდის. საქართველოს საგადასახადო კოდექსის თანახმად, საქართველოს ტერიტორიაზე მოქმედი ყველა ფიზიკური თუ იურიდიული პირი ვალდებულია გადასახადების გადახდის გზით მონაწილეობა მიიღოს სახელმწიფო ბიუჯეტისა და არასაბიუჯეტო ფონდების ფორმირებაში. „ვალდებულება ნიშნავს, რომ ყველა იურიდიულმა და ფიზიკურმა პირმა, რომელიც შემოსავლებს ღებულობს, გააჩნია ქონება და ახორციელებს სამეურნეო ოპერაციებს, აუცილებელია მონაწილეობა მიიღოს სახელმწიფო რესურსების ფორმირებაში" (ფუტკარაძე, 2012).

თეორიულად, საგადასახადო დაბეგვრის სისტემა უნდა შეესაბამებოდეს ეკონომიკური განვითარების იმ მდგომარეობას, რომელშიც ქვეყანა მოცემულ კონკრეტულ სიტუაციაში იმყოფება. ეს იმას ნიშნავს რომ, წარმოების აღმავლობის პირობებში იზრდება საგადასახადო ბაზა, რაც არ გამორიცხავს საგადასახადო განაკვეთების ზრდას, ხოლო კრიზისულ სიტუაციებში როცა წარმოების დონე ეცემა, გადასახადების განაკვეთების შემცირება და მწარმოებლებისთვის საგადასახადო შეღავათების შემოღება, რიგ შემთხვევაში კი რეგრესული განაკვეთების დადგენა ხდება. თუმცა, ეს მხოლოდ თეორიული დასაბუთებაა, რომელსაც შესაძლოა პრაქტიკული გამოყენება არ ჰქონდეს.

საგადასახადო პოლიტიკის რეალიზება საკანონმდებლო და სხვა ნორმატიული აქტებით ხორციელდება. საქართველოში საგადასახადო კანონმდებლობა შედგება საქართველოს კონსტიტუციის, საერთაშორისო ხელშეკრულებებისა და შეთანხმებების და მათ შესაბამისად მიღებული ნორმატიული აქტებისგან, რომლებიც განსაზღვრავს, გადასახადების სახეებს, გადამხდელთა წრეს, განაკვეთებს, საგადასახადო შეღავათებს, გადამხდელის უფლება-მოვალეობებს, პასუხისმგებლობის ზომებს და ა.შ.

ნებისმიერი სახელმწიფოს საგადასახადო პოლიტიკის ჩამოყალიბების საფუძველს ძირითადად, ქვეყნის საზოგადოებრივ-პოლიტიკური მდგომარეობა, საზოგადოების სოციალურ-ეკონომიკური წყობა, ხელისუფლებაში მყოფი სოციალური ჯგუფების ინტერესები და საგადასახადო სამართლებრივ ურთიერთობათა სისტემაში ჩამოყალიბებული ტრადიციები განსაზღვრავენ.

ზოგადად, საგადასახადო პოლიტიკის სამი ტიპი გამოიყოფა: პირველი ტიპი - მაქსიმალური საგადასახადო პოლიტიკა, რაც ძირითადად ორიენტირებულია საგადასახადო განაკვეთების ზრდისკენ. მეორე ტიპი - გონივრული საგადასახადო პოლიტიკა რომელიც, ხელშემწყობი და ხელსაყრელი საგადასახადო კლიმატით მეწარმეობის განვითარებას უზრუნველყოფს. რაც შეეხება მესამე ტიპს, იგი დაბეგვრის შედარებით მაღალ დონეს ითვალისწინებს მნიშვნელოვანი სოციალური დაცვის პირობებში . მძლავრი ეკონომიკის პირობებში საგადასახადო პოლიტიკის სამივე ტიპი წარმატებით შეერწყმის ერთმანეთს, მაგრამ ჩვენი ქვეყნისთვის მეტწილად საგადასახადო პოლიტიკის პირველი ტიპი გამოიყენება მესამესთან ერთად.

უნდა აღინიშნოს, რომ ეფექტიანი საგადასახადო პოლიტიკის გატარებას შეუძლია მნიშვნელოვანწილად შეუწყოს ხელი სახელმწიფოს ძირითადი ეკონომიკური ამოცანების გადაჭრას. საგადასახადო კანონმდებლობაში შესული ცვლილებები სხვადასხვაგვარად მოქმედებს ქვეყანაში მიმდინარე ეკონომიკურ პროცესებზე. ნებისმიერი სახელმწიფოს აღმასრულებელი ხელისუფლების ერთ-ერთ უმნიშვნელოვანეს პრობლემას გამართული და სტაბილური საგადასახადო პოლიტიკის შემუშავება და განხორციელება წამოადგენს, რომელმაც უნდა განსაზღვროს თუ როგორი საგადასახადო-საბიუჯეტო სისტემა და პოლიტიკა შეუწყობს ხელს მოსახლეობის კეთილდღეობის დონის ამაღლებას და მაკროეკონომიკური ამოცანების წარმატებით გადაწყვეტას.

მოქმედ საგადასახადო კანონმდებლობაში კვლავ შეინიშნება ე.წ. „თეთრი ლაქები", რომლებიც გადასახადის გადამხდელებს შესაძლებლობას აძლევს კანონმდებლობაში არსებული დებულებები მათი ხელოვნურად ინტერპრეტირებით სულ სხვაგვარად წარმოადგინონ და მათი დასჯის, დამატებითი გადასახადებისა ან/და ფინანსური სანქციების დაკისრების თავიდან აცილების საფუძვლად გამოიყენონ. საგადასახადო მექანიზმის გამოყენების აუცილებლობა იმითაცაა განპირობებული, რომ ეკონომიკაში შექმნილი დოვლათი თანაბრად იქნეს განაწილებული საზოგადოების ყველა წევრზე. ქვეყანაში შექმნილი ეკონომიკური დოვლათის გადანაწილების უმთავრეს იარაღს საგადასახადო დაბეგვრა წარმოადგენს. სახელმწიფო გადასახადების დაწესებით ბიუჯეტში შემოსავლების მობილიზებას ახორციელებს, რათა ქვეყნის ხელისუფლებამ სახელმწიფოს წინაშე არსებული ფუნქციების შესრულებასთან დაკავშირებული ხარჯები დაფაროს. სახელმწიფოს შემოსავლები და ხარჯები, მათი განაწილება წყაროებისა და მიმართულებების შესაბამისად სახელმწიფო ბიუჯეტის სახით წარმოგვიდგება.

საგადასახადო სისტემის ფორმირების დროს დილემის წინაშე დგას ქვეყნის აღმასრულებელი ხელისუფლება. ამ წინააღმდეგობის ერთ მხარესაა სახელმწიფო ბიუჯეტის შემოსავლების, ხოლო მეორე მხარეს სამეწარმეო აქტიურობის ზრდის საკითხი. შესაბამისად, ჩნდება კითხვა: როგორ უზრუნველვყოთ საბიუჯეტო შემოსავლების გადიდება სამეწარმეო საქმიანობის სტიმულირების პარალელურად?

საგადასახადო პოლიტიკის განხორციელება საზოგადოების გარკვეული ნაწილისთვის შეიძლება ეფექტიანი იყოს, ნაწილისთვის კი საზიანო, დამღუპველიც და უარყოფითიც. და, ეს საბაზრო ეკონომიკისათვის ჩვეული მოვლენაა, რადგან პრაქტიკულად სახელმწიფო ვერ შეძლებს ისეთი საგადასახადო პოლიტიკის შემუშავებას, რომ საზოგადოების ყველა წევრი, მითუმეტეს გადასახადების გადამხდელები, რომლებსაც აქვთ ძირითადი ვალდებულება მათი შემოსავლების სახელმწიფო ბიუჯეტში განაწილებისა, კმაყოფილნი იყვნენ.

ზოგადად, საგადასახადო პოლიტიკა სახელმწიფოს ეკონომიკური პოლიტიკის შემადგენელი ნაწილია, ამიტომაც საქართველოს საგადასახადო სისტემის სრულყოფისათვის აუცილებელია საგადასახადო რეფორმების კომპლექსურად გატარება. უნდა აღინიშნოს ისიც, რომ ქვეყანაში კვლავაც არ არის სრულყოფილად ჩამოყალიბებული ერთიანი ეკონომიკური პოლიტიკა და ამის გამო საქართველოს საგადასახადო კოდექსი მუდმივ კორექტირებას საჭიროებს.

გამომდინარე იქიდან, რომ საბაზრო ეკონომიკა არ არის ჩაკეტილი სისტემა და ნებისმიერი ეკონომიკური სუბიექტის ქმედება გავლენას ახდენს სხვადასხვა ეკონომიკურ გადაწყვეტილებაზე, აუცილებელია საგადასახადო პოლიტიკის ფორმირებისათვის გათვალისწინებული იყოს მასზე მოქმედი ყველა ფაქტორის გავლენა, მოხდეს საგადასახადო სისტემის კომპლექსური ანალიზი და გადაიდგას ისეთი ნაბიჯები რომელიც რეალურად სასარგებლო და პერსპექტიული იქნება ჩვენი ქვეყნისათვის.

## 1.2 პირდაპირი და არაპირდაპირი გადასახადების როლი და ადგილი ბიუჯეტის საგადასახადო შემოსავლებში

საგადასახადო შემოსავლების სტრუქტურაში არსებითი მნიშვნელობა პირდაპირი და არაპირდაპირი გადასახადების თანაფარდობას ენიჭება. საბაზრო ეკონომიკის ქვეყნების პრაქტიკა ადასტურებს, რომ თანაფარდობა პირდაპირ და არაპირდაპირ გადასახადებს შორის დამოკიდებულია ქვეყნის მოსახლეობის ძირითადი ნაწილის ცხოვრების დონეზე, მოსახლეობის შემოსავლების საშუალო დონეზე და გადასახადების დარიცხვისა და ამოღების მეთოდურ-სამართლებრივ საფუძვლებზე.

პირდაპირი გადასახადები უშუალოდ, შემოსავლების მიღების პროცესში ამოიღება, ხოლო არაპირდაპირი - მიღებული შემოსავლების ხარჯვისას. პირდაპირი გადასახადები საგადასახადო სისტემის საფუძველს წარმოადგენს, იგი ქვეყანაში საგადასახადო წნეხის სამართლიანად გადანაწილებას უწყობს ხელს. მას მიეკუთვნება მოგების, საშემოსავლო და ქონების გადასახადი, ხოლო არაპირდაპირს - დამატებული ღირებულების გადასახადი, აქციზი და იმპორტის გადასახადი.

არაპირდაპირი გადასახადები განისაზღვრება მიწოდებული საქონლის გაყიდული მომსახურების ან შესრულებული სამუშაოს ფასთან მიმართებით. არაპირდაპირი გადასახადი საგადასახადო კოდექსში განმარტებულია როგორც: „გადასახადი (დამატებული ღირებულების გადასახადი, აქციზი, იმპორტი), რომელიც დგინდება მიწოდებული საქონლის ან/და გაწეული მომსახურების ფასზე დანამატის სახით და რომელსაც იხდის მომხმარებელი ამ გადასახადით გაზრდილი ფასით საქონლის ან/და მომსახურების შეძენისას. არაპირდაპირი გადასახადის ბიუჯეტში გადახდის ვალდებულება ეკისრება საქონლის მიმწოდებელს ან/და მომსახურების გამწევს, რომელიც ამ კოდექსის მიზნებისათვის იწოდება გადასახადის გადამხდელად" (საქართველოს საგადასახადო კოდექსი, 2010) თანამედროვე პირობებში არაპირდაპირი გადასახადები წარმოადგენს ქმედით ინსტრუმენტს საგადასახადო წნეხის მწარმოებლიდან მომხმარებელზე გადაკისრებაში. აღნიშნულიდან გამომდინარე, არაპირდაპირი გადასახადების გადამხდელია უშუალოდ საქონლისა და მომსახურების საბოლოო მომხმარებელი და ამ გადასახადების მთელი სიმძიმე მათ აწევთ; საგადასახადო შემოსავლების უდიდესი ნაწილი არაპირდაპირ გადასახადებზე მოდის. ამიტომაც მნიშვნელოვანია განისაზღვროს საგადასახადო შემოსავლებში არაპირდაპირი გადასახადების წილი, მათი როლი და მნიშვნელობა.

ქვემოთ წარმოდგენილ ცხრილში მოცემულია 2014-2019 წლების საგადასახდო შემოსავლებში არაპირდაპირი გადასახადებით მობილიზებული ფინანსური რესურსების ხვედრითი წილი. საგადასახადო შემოსავლების დინამიკაში ანალიზი დაგვეხმარება განვსაზღვროთ მათი ზრდის და კლების ტენდენციები, გამომწვევი მიზეზები თუ მოქმედი ფაქტორები. გამომდინარე იქედან, რომ ეკონომიკაში შემთხვევით არაფერი ხდება, საბაზრო ეკონომიკა არ არის ჩაკეტილი სივრცე და ერთი ეკონომიკური გადაწყვეტილების, მოვლენისა თუ პროცესის ცვლილება პირდაპირ კავშირს ახდენს სხვა მაკროეკონომიკური პარამეტრების ცვლილებაზე; ამიტომაც ზუსტად უნდა განვსაზღვროთ და გავაანალიზოთ საგადასახდო შემოსავლების თუნდაც ერთი ერთეულით გაზრდის თუ შემცირების მიზეზები, მოქმედი ფაქტორები და მოსალოდნელი შედეგები სხვა მაკროეკონომიკურ ცვლადებზე.

ცხრილი 2. არაპირდაპირი გადასახადების წილი სახელმწიფო ბიუჯეტის საგადასახადო შემოსავლებში 2014-2019 წლებში (ათასი ლარი)

|  | 2014 | 2015 | 2016 | 2017 | 2018 | 2019* |
|---|---|---|---|---|---|---|
| საგადასახადო შემოსავლები | 6,846,964.2 | 7,549,608.9 | 7,986,750.3 | 8,991,307.5 | 9,695,962.2 | 9,645,000.0 |
| დამატებული ღირებულების გადასახადი | 3,298,518.3 | 3,505,454.6 | 3,286,393.4 | 4,122,612.8 | 4,426,909.8 | 4,036,000.0 |
| აქციზი | 810,209.1 | 870,731.1 | 1,069,658.1 | 1,450,921.9 | 1,465,726.6 | 1,462,000.0 |
| იმპორტი | 94,883.2 | 69,294.0 | 70,040.6 | 71,618.9 | 73,416.9 | 83,000.0 |
| %-ული წილი საგადასახადო შემოსავლებში | 61,39 | 58,88 | 55,42 | 62,78 | 61,53 | 57,86 |

წყარო: შედგენილია ავტორის მიერ საქართველოს ფინანსთა სამინისტროს მონაცემებზე დაყრდნობით როგორც №1 და №2 ცხრილიდან ჩანს საგადასახადო შემოსავლების უდიდესი ნაწილი არაპირდაპირ გადასახადებზე მოდის. 2014-2017 წლებში აღნიშნული გადასახადების წილი დაახლოებით 61-63%-ის ფარგლებში მერყეობდა, 2018 წელს იგი მცირედით შემცირდა (61,53%_მდე) (უნდა აღინიშნოს ისიც, რომ 2019 წლების მონაცემები არ არის დაზუსტებული).

საერთაშორისო პრაქტიკა ადასტურებს, რომ თუ სახელმწიფო ორიენტირებულია არაპირდაპირ გადასახადებზე, მაშინ, საქმე გვაქვს დაბალ განვითარებულ ქვეყანასთან, ხოლო თუ გადასახადების სტრუქტურაში ჭარბობს პირდაპირი გადასახადები, მაშინ ეს მიუთითებს, რომ საუბარია მაღალგანვითარებულ ეკონომიკაზე. ეს კანონზომიერება თვალნათლივ წარმოჩინდება განვითარებად ქვეყნებში, რომელთა საგადასახადო დაბეგვრაში დომინირებს არაპირდაპირი გადასახადები.

არაპირდაპირ გადასახადებს შორის ყველაზე მაღალი წილი დამატებული ღირებულების გადასახადს უჭირავს, ამიტომაც ჩვენი ყურადღება ძირითადად დამატებული ღირებულების გადასახადზე მიმართული.

დამატებული ღირებულების გადასახადი ირიბ, მრავალსაფეხურიან გადასახადს წარმოადგენს, რომლის ძირითადი ფუნქციაც ფისკალურია. დამატებული ღირებულების გადასახადიდან შემოსული მთლიანი თანხა საქართველოს სახელმწიფო ბიუჯეტში მიემართება. იგი ქვეყნის ერთ-ერთი ძირითადი გადასახადია და ცენტრალური ბიუჯეტის საშემოსავლო ნაწილის ფორმირებაში მნიშვნელოვან როლს ასრულებს. დამატებული ღირებულების გადასახადი წარმოებისა და მიმოქცევის სტადიაზე შექმნილი დამატებული ღირებულების ნაწილის ბიუჯეტში ამოღების ფორმას წარმოადგენს. დამატებული ღირებულების გადასახადით იბეგრება არა მარტო მიმოქცევა, არამედ წარმოების მთელი სტადია, ანუ წარმოებისა და საქონელ მოძრაობის ჯაჭვში არსებული რგოლიდან ყოველი მონაწილე დამატებული ღირებულების გადასახადის გადახდაში მონაწილეობს.

დამატებული ღირებულების გადასახადის დინამიკას (ცხრილი №2) თუ დავაკვირდებით 2016 წლიდან „ნახტომისებური" ზრდა შეინიშნება. თითქმის ერთი მილიონითა გაზრდილი 2017 წლისათვის დამატებული ღირებულების გადასახადით მოზიდული სახსრები წინა წელთან შედარებით. 2017 წელს დამატებული ღირებულების გადასახადის სახით მობილიზებული (4 122 612.8) რესურსები საპროგნოზო მაჩვენებლის (4 020 000.0 ათასი ლარი) 102.6%-ს შეადგენს. აღსანიშნავია, რომ 2017 წელს წინა წელთან შედარებით შესამჩნევად გაიზარდა იმპორტის მაჩვენებელი ჩვენს ქვეყანაში (7,293.8-7,939.1 მლნ. ლარი), რამაც პირდაპირი ასახვა ჰპოვა დამატებული ღირებულების გადასახადზეც, გამომდინარე იქიდან, რომ დამატებული ღირებულების გადასახადი იმპორტირებულ საქონელს უკვე დარიცხული გადასახადების შემდგომ ერიცხება;

იმპორტის წინმსწრები მატება მნიშვნელოვან გავლენას ახდენს ბიუჯეტის საგადასახადო შემოსავლებზე. უშუალოდ იმპორტის გადასახადის დინამიკას თუ დავაკვირდებით იგი იმავე ტემპით იცვლება რა ტემპითაც იცვლება იმპორტის რაოდენობა ჩვენს ქვეყანაში. გამომდინარე იქედან, რომ ჩვენი ქვეყანა ძირითადად იმპორტირებულ პროდუქციაზეა დამოკიდებული, იმპორტის გადასახადს ერთ-ერთი მნიშვნელოვანი ადგილი უჭირავს საბიუჯეტო შემოსავლებში. ბოლო 5 წლის განმავლობაში იმპორტის გადასახადით მოზიდული სახსრები აბსოლუტური ზრდის ტენდენციით გამოირჩევა, რაც დადებითი ფაქტორია საგადასახადო შემოსავლებისთვის, მაგრამ იგივეს ვერ ვიტყვით, ადგილობრივ წარმოებასა და ქვეყანაში ჯანსაღი კონკურენტული გარემოს ჩამოყალიბებაზე.

აქციზის გადასახადით მობილიზებული სახსრები, ბოლო წლების განმავლობაში ასევე ზრდის ტემპით ხასიათდება, რადგანაც აქციზის გადასახადი საგადასახადო დაბეგვრაში ასრულებს, როგორც ფისკალურ ასევე მარეგულირებელ ფუნქციასაც და შესაბამისად, იგი წარმოადგენს ერთგვარ ინსტრუმენტს გარკვეული ჯგუფის საქონლის მოხმარების შემცირებაში და საგადასახადო შემოსავლების ზრდაში. თუმცა აღნიშნული გადასახადის განაკვეთსა და პროდუქციაზე მოთხოვნას შორის ურთიერთდამოკიდებულება, ჩვენი ქვეყნის მაგალითზე წმინდა ეკონომიკურ თეორიებს სცილდება, რადგანაც გადასახადის განაკვეთის ზრდა აქციზურ საქონელზე მოხმარებას არსებითად არ ამცირებს, გამომდინარე იქედან, რომ საზოგადოების უმეტესი ნაწილი „მავნე" პროდუქტების (ალკოჰოლი, სიგარეტი და ა.შ) მოხმარებაზე მიდრეკილი.

რაც შეეხება მომდევნო წლის მაჩვენებლებს, აღნიშნული მონაცემები საპროგნოზოა და ჯერ მათი ანალიზისთვის გამოყენება ნაადრევია, თუმცა ხაზი უნდა გაესვას იმ ფაქტს, რომ 2018 წელს მიღებული იქნა ახალი რეფორმა მცირე ბიზნესის სტატუსის ფლობასთან დაკავშირებით, რომლის თანახმადაც მცირე ბიზნესის სტატუსი მიენიჭება მეწარმე ფიზიკურ პირებს, რომელთა ეკონომიკური საქმიანობიდან მიღებული შემოსავალი კალენდარული წლის განმავლობაში 500 000 ლარს არ აღემატება, რაც იმას ნიშნავს, რომ 2018 წლიდან მცირე ბიზნესის სტატუსით მოსარგებლე მეწარმეები, რომელთა ეკონომიკური საქმიანობიდან მიღებული შემოსავალი კალენდარული წლის განმავლობაში 100 000 ლარიდან 500 000 ლარამდეა სავალდებულო წესით დამატებული ღირებულების გადასახადი გადამხდელად უნდა დარეგისტრირდნენ, შედეგად გაიზრდება დამატებული ღირებულების გადასახადის გადამხდელთა რაოდენობაც და აღნიშნული გადასახადით მოზიდული სახსრებიც. ზოგადად, არაპირდაპირი გადასახადები სოციალურად ყველაზე არასამართლიანია. ისინი თანაბრად ზემოქმედებენ საზოგადოების ყველა წევრზე. ამასთან, მათი ზეწოლა განსაკუთრებით უარყოფით გავლენას ახდენს მოსახლეობის დაბალშემოსავლიან ჯგუფებზე. გამომდინარე აქედან, სოციალური სამართლიანობის მიღწევის მიზნით, საგადასახადო სისტემამ უნდა დაუშვას ცალკეულ შეღავათები მოსახლეობის დაბალ შემოსავლიანი ჯგუფებისათვის. არაპირდაპირი გადასახადების განხილვისას ძირითადი აქცენტი უნდა გაკეთდეს მომხმარებელზე ანუ ამ გადასახადის საბოლოო გადამხდელზე.

საგადასახადო კოდექსის მიხედვით გადასახადების გადახდის პირდაპირი ვალდებულება მეწარმე ფიზიკურ და იურიდიულ პირებს ეკისრებათ, მაგრამ პრაქტიკულად ეს ვალდებულება მთლიანად ფიზიკურ პირებზე გადადის (გადაკისრების მექანიზმის თანახმად), რომელთაც მათი გადახდა თავიანთი პირადი შემოსავლებიდან უწევთ. იურიდიულ პირებს პრაქტიკულად მოგების გადასახადის გადახდის ვალდებულება "შემორჩათ", ეს უკანასკნელიც კი 2017 წლის 1-ლი იანვრიდან შემოღებული "ესტონური მოდელის" დანერგვით "წაერთვათ", რომელიც გადასახადის გადახდას მხოლოდ მოგების განაწილების შემთხვევაში ითვალისწინებს.

პირდაპირი და არაპირდაპირი გადასახადების სტრუქტურა ყველა ქვეყანაში განსხვავებულია, რაც დამოკიდებულია ქვეყნის ეკონომიკურ მდგომარეობაზე, უშუალოდ საგადასახადო პოლიტიკის ფორმირებასა და რეალიზაციაზე. რაც შეეხება ჩვენი ქვეყნისათვის, არანაკლებ მნიშვნელოვანი ადგილი აქვს ბიუჯეტის საგადასახადო შემოსავლში პირდაპირ გადასახადებს, ამიტომაც აუცილებელი მათი დინამიკაში განხილვა. ცხრილი №3-ზე ასახულია 2014-2019 წლებში პირდაპირი გადასახადებით მოზიდული სახსრების ხვედრითი წილი. როგორც ვიცით, ქონების გადასახადიც პირდაპირი გადასახადის სახეა მაგრამ, რადგანაც ქონების გადასახადით მოზიდული სახსრები ადგილობრივ ბიუჯეტში რჩება მის გამოყენებას ანალიზისა და საერთო სურათის დასანახად არსებითი მნიშვნელობა არ აქვს.

**ცხრილი 3.** პირდაპირი გადასახადების წილი სახელმწიფო ბიუჯეტის საგადასახადო შემოსავლებში 2014-2019 წლებში (ათასი ლარი)

|  | 2014 | 2015 | 2016 | 2017 | 2018 | 2019* |
|---|---|---|---|---|---|---|
| საგადასახადო შემოსავლები | 6,846,964.2 | 7,549,608.9 | 7,986,750.3 | 8,991,307.5 | 9,695,962.2 | 9,645,000.0 |
| საშემოსავლო გადასახადი | 1,790,379.5 | 2,052,388.5 | 1,978,136.4 | 2,525,97.7 | 2,877,895.1 | 3,166,000.0 |
| მოგების გადასახადი | 828,823.2 | 1,025,228.5 | 1,055,936.5 | 756,555.5 | 736,624.4 | 781,000.0 |
| %-ული წილი საგადასახადო შემოსავლებში | 38,25 | 40,77 | 37,99 | 36,51 | 35,93 | 40,92 |

**წყარო:** შედგენილია ავტორის მიერ საქართველოს ფინანსთა სამინისტროს მონაცემებზე დაყრდნობით საშემოსავლო გადასახადი ძირითადად ზრდის ტენდენციით ხასიათდება რაც მისასალმებელი ფაქტია, 2019 წლისათვის კი 3 მილიონ ლარზე მეტი რესურსის მიღებაა პროგნოზირებული.

საშემოსავლო გადასახადით დაბეგვრის წესი და ნორმები ყოველთვის ფართო ინტერესის საგანია დამსაქმებელთა, დასაქმებულთა და სახელმწიფოს შორის. საქართველოში აღნიშნულმა გადასახადმა მრავალჯერადი ცვლილება განიცადა. 2007 წლიდან მოყოლებული საშემოსავლო გადასახადმა მკვეთრი ტრანსფორმაცია განიცადა და დღეს 20%-იანი განაკვეთით არის წარმოდგენილი, გამონაკლის წარმოადგენს დივიდენტისა და პროცენტის სახით მიღებული შემოსავალი რომელიც 5%-იანი განაკვეთით იბეგრება. საერთო ჯამში, საშემოსავლო გადასახადებით მოზიდული სახსრები ბიუჯეტის საგადასახადო შემოსავლების 25%-28% შეადგენს (ცხრილი №2).

რაც შეეხება მოგების გადასახადის დინამიკას, დაახლოებით 300 000 ლარითაა შემცირებული მოგების გადასახადით მობილიზებული რესურსები (2017) წინა წელთან შედარებით, რისი უმთავრესი გამომწვევი მიზეზიც 2017 წელს მოგების ე.წ „ესტონური" მოდელით დაბეგვრის რეჟიმის შემოღებაა, რომელიც მოგების გადასახადით დაბეგვრას მხოლოდ მოგების განაწილებისას ითვალისწინებს. გაუქმდა პროპორციული 15%-იანი განაკვეთი და განისაზღვრა ე.წ „აგროსვითი" გადასახადის განაკვეთი (15/85); უშუალოდ „ესტონური" მოდელის შემოღებით გამოწვეულ შედეგებზე შემდეგ თავში ვისაუბრებთ, მაგრამ ფაქტი ის არის, რომ ახალი რეფორმა უარყოფითად აისახება საგადასახადო შემოსავლებზე.

მოგებაზე გადასახადი პირდაპირი გადასახადია, რომელიც რეზიდენტობის პრინციპზეა დაფუძნებული. მისი ძირითადი გადამხდელი მეწარმე იურიდიული პირები არიან, რომლებიც კანონმდებლობის შესაბამისად ეკონომიკურ საქმიანობას ეწევიან. საგადასახადო კანონმდებლობის ცვალებადობას თუ დავაკვირდებით, გამუდმებით მიმდინარეობდა მუშაობა მოგების გადასახადის ამოღების ოპტიმიზაციისათვის, კერძოდ, აღნიშნული გადასახადით დაბეგვრის ობიექტის სწორად განსაზღვრის, საგადასახადო განაკვეთების შემცირებისა და საგადასახადო შეღავათების მოქნილად და რაციონალურად გამოყენებისთვის.

უკვე არა ერთხელ აღინიშნა, რომ არაპირდაპირი გადასახადების უშუალო გადამხდელია წარმოებული პროდუქციისა და მომსახურების საბოლოო მომხმარებელი, რომელთა მიერ გადახდილ ფასში უკვე დარიცხულია გარკვეული სახის არაპირდაპირი გადასახადები. ამ პროდუქციის მწარმოებელი კი მხოლოდ საგადასახადო აგენტის როლს ასრულებს გადასახადის ბიუჯეტში გადახდისას. ასეთ პროდუქციასა და მომსახურებას, რომელთა ფასში უკვე შესულია არაპირდაპირი გადასახადები, დღის განმავლობაში ასობით და ათასობით ადამიანი ყიდულობს, იქნება ეს სოციალურად მაღალი ფენის წარმომადგენელი თუ სახელმწიფოსგან საარსებო შემწეობის მომლოდინე პირი. ბუნებრივია ამ ორი რადიკალურად განსხვავებული კატეგორიის გადამხდელს განსხვავებული სიმძიმით აწვება აღნიშნული გადასახადი. ამიტომაც აღნიშნული პრობლემის აღმოსაფხვრელად, ვთვლით რომ ეფექტიანი იქნება მოხდეს არაპირდაპირი გადასახადების განაკვეთების დიფერენციაცია პროდუქციის სახეების მიხედვით.

აღნიშნული ცვლილება გულისხმობს, რომ ისეთი პირველადი მოხმარების საქონელზე, როგორიცაა მაგალითად: პური, ზეთი, შაქარი, მარილი და ა.შ შემცირდეს არაპირდაპირი გადასახადების დადგენილი განაკვეთი. ეს დიფერენციაცია გარკვეულწილად უკვე არსებობს აქციზის და იმპორტის გადასახადის შემთხვევაში და გარკვეული სახეობის საქონელზე განსხვავებული ტარიფებია შემოდებული, მაგრამ მსგავსი რეფორმის დამატებული ღირებულების გადასახადისთვისაც შემოდება უკეთესი იქნება. ვთვლით, რომ აღნიშნული რეფორმა სასარგებლო იქნება სოციალურად დაბალი ფენისათვის, რომელთა წილი საზოგადოებაში გაცილებით უფრო მაღალია, ვიდრე საქართველოს სტატისტიკის ეროვნული სამსახური გვატყობინებს (2017 წლის მაჩვენებელი - 21,9%).

აღნიშნული ცვლილება რა თქმა უნდა უარყოფითად აისახება ბიუჯეტის საგადასახადო შემოსავლებზე, რომლის ძირითად საშემოსავლო ნაწილს სწორედ დამატებული ღირებულების გადასახადი წარმოადგენს, მაგრამ მაშინ როდესაც ვამბობთ, რომ საქართველოს სახელმწიფო ბიუჯეტი სოციალურად ორიენტირებულია, ვფიქრობ სახელმწიფო უფრო მეტად უნდა ცდილობდეს საზოგადოების დაბალ შემოსავლიანი ფენის მატერიალურ კეთილდღეობაზე. ამიტომაც ვამბობთ, რომ არაპირდაპირი გადასახადი სოციალურად არასამართლიანია, როცა პურის და ძვირადღირებული ავტომობილის შეძენისას თანაბარი 18%-იანი დამატებული ღირებულების გადასახადის განაკვეთი გამოიყენება. ამ ფაქტს ხელისუფლების მხრიდან ყურადღება და ღრმა ანალიზი სჭირდება.

დიაგრამა №2 ნათელ სურათს გვაძლევს იმის შესახებ თუ რა პროპორციითა განაწილებული ბიუჯეტის საგადასახადო შემოსავლებში პირდაპირი და არაპირდაპირი გადასახადებით მობილიზებული რესურსები.

დიაგრამა 2. 2014-2019 წლების სახელმწიფო ბიუჯეტის შემოსავლებში პირდაპირი და არაპირდაპირი გადასახადები (ათასი ლარი)

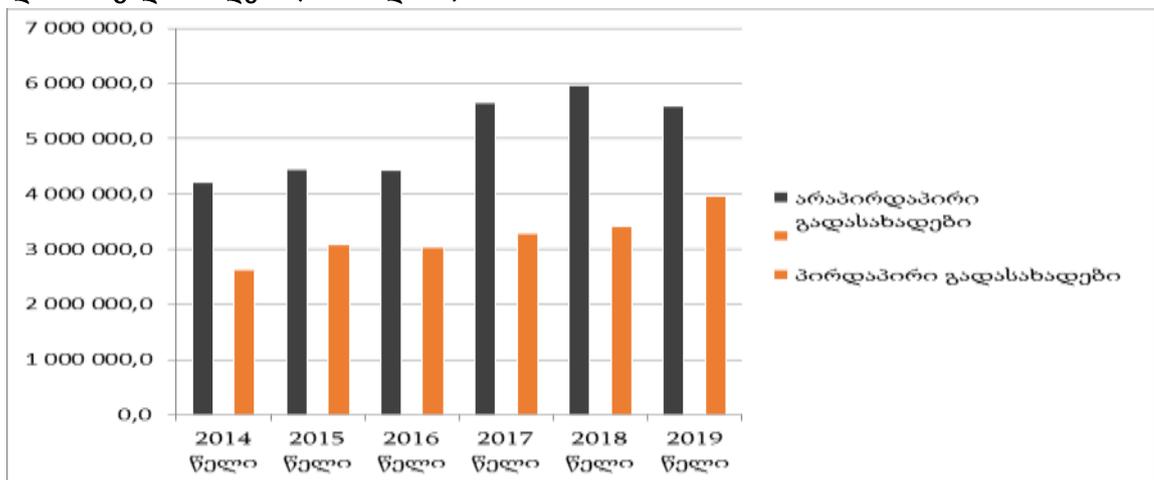

წყარო: შედგენილია ავტორის მიერ საქართველოს ფინანსთა სამინისტროს მონაცემებზე დაყრდნობით პირდაპირი და ირიბი გადასახდების როლის ამაღლებისთვის აუცილებელია, უპირველესად, სამეწარმეო სფეროში შეიქმნას კეთილსასურველი საგადასახადო გარემო, გაიზარდოს ფიზიკურ პირთა და საწარმოთა რეალური შემოსავლები, ამაღლდეს საგადასახადო კულტურა, რომელიც მოქმედებს გადასახადების ამოღების ეფექტურობაზე, სრულყოფილი გახდეს საგადასახადო კანონმდებლობა და შემოსავლების სამსახურების საქმიანობა. საგადასახადო კანონმდებლობაში უმნიშვნელოვანესია მისი ორმაგად გაგების პრობლემის გამოსწორება, რომელიც „მავანს და მავანს" საშუალებას აძლევს იგი მათ სასარგებლოდ გამოიყენონ, დაფარონ მათი რეალური შემოსავლები ან/და თავი აარიდონ საგადასახადო ვალდებულებების შესრულებას. სრულყოფილი საგადასახადო კლიმატის შესაქმნელად აუცილებელია საგადასახადო ვალდებულებების შესრულებაზე კონტროლის მექანიზმების განხორციელება, რაც გულისხმობს გარკვეული სპეციფიკური პროცედურების განხორციელებას სამართლებრივი, ორგანიზაციული, მატერიალურ-ტექნიკური და სოციალურ-ფსიქოლოგიური მიმართულებით.

## 2. საგადასახადო შემოსავლების ფორმირების სრულყოფის მიმართულებები
### *2.1 საგადასახადო სტიმულირების ღონისძიებები და მისი გავლენა ბიუჯეტის საგადასახადო შემოსავლებზე*

საქართველოს საგადასახადო კანონმდებლობა მუდმივად კორექტირებასა და ცვლილებას განიცდის, მისი გაუმჯობესებისა და ოპტიმიზაციის მიზნით. ჯერ კიდევ 2003 წლიდან ძირეული ცვლილებები შევიდა საგადასახადო კოდექსში და აქამდე არსებული 36 გადასახადი, 6 გადასახადამდე შემცირდა. ამჟამად მოქმედი საგადასახადო კოდექსი 2010 წლის 17 სექტემბერსაა მიღებული და დღემდე მასში 150-ზე მეტი ცვლილებაა შესული. საკანონმდებლო ცვლილებების მიზანია უფრო სრულყოფილი და გამართვებული გახდეს საგადასახადო სისტემა, რომელიც ხელს შეუწყობს ბიზნესის დაწყების და განვითარებისთვის ხელსაყრელი პირობების შექმნას და შესაბამისად, ეკონომიური ზრდის დაჩქარებას. ფრანგ ფილოსოფოს მონტესკიეს აზრით: „საზოგადოებაში არაფერი საჭიროებს იმდენ ჭკუა-გონებას, რამდენიც აუცილებელია იმ წილის განსასაზღვრავად, რომელსაც ხელისუფლება ართმევს საზოგადოების წევრებს", ამიტომაც საგადასახადო კანონმდებლობაში ზუსტად და გასაგებად უნდა იყოს ჩამოყალიბებული საგადასახადო ორგანოსა და გადასახადის გადამხდელის უფლება-მოვალეობები.

აბუსელიძის აზრით ეკონომიკურად ძლიერი სახელმწიფო დაფუძნებულია საგადასახადო სისტემის ეფექტიან ფუნქციონირებაზე, რომლისთვისაც აუცილებელია არა მხოლოდ ოპტიმალური საგადასახადო განაკვეთები, არამედ თითოეული გადამხდელის მაღალი საგადასახადო კულტურა, როგორიცაა:

გადასახადების აკრეფის, მისი გადახდის აუცილებლობის შეგნება, თითოეული მეწარმის და გადასახადის ამკრეფის მაღალი პასუხისმგებლობა და მთლიანად ადმინისტრირების ხარისხი (Abuselidze, 2013).

საგადასახადო რეფორმებს ძირითადად ორი ეფექტი აქვს: ერთის მხრივ, ის შეიძლება უარყოფითად აისახოს ბიზნეს სუბიექტების საქმიანობაზე და გარკვეულ ეტაპზე შეაფერხოს კიდეც, თუმცა შესაძლებელია იგი დამატებითი სტიმული აღმოჩნდეს სამეწარმეო საქმიანობისათვის. იგივე ეფექტი აქვს საგადასახადო რეფორმის გატარებას ბიუჯეტის საგადასახადო შემოსავლებისთვისაც, ამიტომაც „ოქროს შუალედის" მიღწევა ძალიან რთულია, მით უფრო ჩვენი ქვეყნისათვის, რომელიც ჯერ კიდევ ავითარებს საბაზრო ეკონომიკის პრინციპებს. ერთია, ის თუ აღმასრულებელი ხელისუფლება რა მიზნით ატარებს ამა თუ იმ საგადასახადო რეფორმას და მეორეა თუ როგორ ღებულობენ და ითვალისწინებენ მას საგადასახადო სუბიექტები.

ნებისმიერი სახელმწიფოს უმნიშვნელოვანესი პრიორიტეტი საგადასახადო სისტემის ოპტიმიზაცია უნდა იყოს, ვინაიდან მის გამართულ ფუნქციონირებაზე არის დამოკიდებული სახელმწიფოს მიერ აუცილებელი ხარჯების დაფინანსების შესაძლებლობა. ამასთან, ოპტიმიზაციაში უნდა ვიგულისხმოთ ორი უმნიშვნელოვანესი ამოცანის გადაწყვეტა – 1) ბიუჯეტში შემოსავლების მაქსიმალურად მობილიზება და 2) გადამხდელთა გადახდისუნარიანობის გათვალისწინება საგადასახადო განაკვეთების დადგენისას. ამასთან, ყურადღების მიღმა არ უნდა დავგრჩეს გარემოება, რომ საგადასახადო სიმძიმის მთლიანად მისაღებობის პირობებშიც კი, გადასახადებს, მისი არამართლზომიერი განაწილების გამო, შეიძლება მაინც ქონდეს სერიოზული დამახინჯებული სახე (Abuselidze, 2015).

საგადასახადო კოდექსში შეტანილ ცვლილებებს თან სდევს რამდენიმე მნიშვნელოვანი საკითხი, მათ შორის, ის თუ რამდენად ეფექტიანი იქნება აღნიშნული რეფორმა ბიუჯეტის საგადასახადო შემოსავლებისთვის და ზოგადად, ის, თუ როგორ მიიღებს მას საზოგადოება. თუმცა ამ უკანასკნელს უპირატესი მნიშვნელობა არ ენიჭება, რადგანაც საქართველოში საგადასახადო კულტურასთან დაკავშირებით მწვავე პრობლემები გვაქვს, რომელიც ცალკე კვლევის საგანს წარმოადგენს; რატომდაც საქართველოში ითვლება, რომ გადასახადის გადახდის ვალდებულება არავის აქვს და ყველა ცდილობს სხვადასხვაგვარი მეთოდებით თავი აარიდოს მათ. სწორედ ამიტომ, აღმასრულებელი ხელისუფლება ხშირად ატარებს საგადასახადო სტიმულირების ღონისძიებებს, რათა ხელი შეუწყოს სამეწარმეო საქმიანობის ეფექტიანად განხორციელებას, რაც ასევე დადებითად აისახება სახელმწიფო ბიუჯეტის საგადასახადო შემოსავლებზე. უახლოეს წარსულს თუ გადავხედავთ, ბევრი მნიშვნელოვანი ცვლილება შევიდა საგადასახადო კოდექსში. სახელმწიფო მაქსიმალურად ცდილობს სხვადასხვა მასტიმულირებელი ღონისძიებების გატარებით უფრო მარტივი, ნათელი და არაორაზროვანი გახდეს საგადასახადო კანონმდებლობა, რათა გადასახადის გადამხდელმა პირნათლად შეასრულოს საგადასახადო ვალდებულებები. ჯერ კიდევ 2010 წლიდან გადასახადის გადამხდელებისათვის მაქსიმალური კომფორტის უზრუნველსაყოფად შემოსავლების სამსახურმა დანერგა მრავალი სახის ელექტრონული მომსახურება. თანამედროვე ტექნოლოგიების დანერგვა მნიშვნელოვანია გადამხდელზე ორიენტირებული სერვისების შეთავაზებისთვის. ვებ-პორტალის შექმნა შემოსავლების სამსახურის მნიშვნელოვანი მიღწევა იყო. იგი შემოსავლების სამსახურის მრავალრიცხოვანი მომსახურების პორტალია, რომლის მეშვეობითაც მნიშვნელოვნად იზოგება გადამხდელის დრო, ენერგია და ფინანსური რესურსები. მისი მეშვეობით, ადგილიდან გაუსვლელად, ზედმეტი დანახარჯების გარეშე შესაძლებელი გახდა შემოსავლების სამსახურთან კომუნიკაცია. გადასახადების ელექტრონული დეკლარირება, ინტერნეტ-გადახდები, განცხადებებისა და დოკუმენტაციის ელექტრონულად წარდგენა და მათი შესრულების მდგომარეობაზე თვალყურის დევნება უკვე შესაძლებელია ონლაინ რეჟიმში. გარდა ამისა, შემოსავლების სამსახურმა პირადი საგადასახადო აგენტის სახით, მომსახურების ახალი ტიპი დანერგა (შემოსავლების სამსახური, 2010). საგადასახადო აგენტი გადასახადის გადამხდელებს უადვილებს საგადასახადო ორგანოებთან ურთიერთობას, აძლევს რჩევებს, ეხმარება საგადასახადო კანონმდებლობაში შეტანილი სიახლეების დროულად გაცნობაში. თითოეული აგენტი პასუხისმგებელია საკუთარ რეკომენდაციაზე და შესაბამისად, მეწარმე სუბიექტები დაზღვეულები არიან შეცდომებისგან. 2010 წლიდან ასევე ამოქმედდა ახალი რეფორმა, რომელიც მიმართული იყო სახელმწიფოსა და კერძო სექტორს შორის პარტნიორული ურთიერთობის ჩამოყალიბებისკენ. აღნიშნული რეფორმა გულისხმობდა საკონტროლო-სალარო აპარატის გამოუყენებლობის შემთხვევაში ჯარიმის ნაცვლად გაფრთხილების მიცემას. სამართალდარღვევის დაფიქსირების შემთხვევაში მეწარმეს უნდა განემარტოს, რომ აღნიშნული წარმოადგენს სამართალდარღვევას და შემოსავლების სამსახური ფულად ჯარიმას მხოლოდ განმეორების შემთხვევაში გამოიყენებს. ჯარიმის ნაცვლად გაფრთხილების შემოღება, კიდევ ერთხელ უსვამს ხაზს მთავრობის ახალ კურსსა და ახლებურ მიდგომას კერძო სექტორისადმი. „სახელმწიფოს სურს ყველა მეწარმეში დაინახოს საგადასახადო კანონმდებლობის დამცველი და პატიოსანი მეწარმე, რომელიც განზრახ კი არ არღვევს კანონმდებლობით დაკისრებულ მოვალეობებს, არამედ არცოდნის ან არასათანადო ინფორმირებულობის გამო ხდება სამართალდამრღვევი" (შემოსავლების სამსახური, 2010). ნაცვლად ფულადი ჯარიმისა მეწარმის ინფორმირება და გაფრთხილება გაზრდის ურთიერთნდობას და პატივისცემას საგადასახადო ორგანოების მიმართ, ანუ ყოველ შემთხვევაში ამ რეფორმის ავტორებს ასეთი მოლოდინი აქვთ, თუმცა ფაქტია, რომ 2010 წელს გაფრთხილება 164 მეწარმეს მიეცა და შედეგად მათ 74 200 ლარის ოდენობის ჯარიმა არ დაერიცხათ. ამოქმედდა კიდევ ერთი საკანონმდებლო მექანიზმი - საგადასახადო შეთანხმება. „გადასახადის გადამხდელს მიენიჭა უფლება, მიმართოს შემოსავლების სამსახურს გადასახადების ან/და სანქციების შემცირების მიზნით და განსაზღვროს ის საგადასახადო პერიოდი ან პერიოდები, რომლის მიმართაც სურს, გამოიყენოს საგადასახადო შეთანხმების მექანიზმი" (შემოსავლების სამსახური, 2010). საგადასახადო შეთანხმება ასევე შესაძლებელია გაფორმდეს გადასახადების სახეების მიხედვით. საგადასახადო შეთანხმების რეფორმა წარმოადგენს კიდევ ერთ საკანონმდებლო ინსტრუმენტს, რომლის დახმარებითაც გადასახადის გადამხდელი შეძლებს პირნათლად შეასრულოს დაკისრებული საგადასახადო ვალდებულება, რაც შემდგომში სახელმწიფოსა და კერძო ბიზნესს შორის პარტნიორული და ნდობაზე დამყარებული ურთიერთობების საფუძველს წარმოადგენს. საგადასახადო შეთანხმების მექანიზმის შემოღებამ უფრო სამართლიანი გახადა გადამხდელზე დაკისრებული პასუხისმგებლობა, რადგანაც გადამხდელზე დარიცხული ჯარიმის მოცულობა ყოველთვის არ არის ადეკვატური განხორციელებული სამართალდარღვევისა და დარიცხულმა თანხამ შესაძლებელია სამეწარმეო საქმიანობის შეფერხება და გარკვეულ ეტაპზე შეჩერებაც კი გამოიწვიოს, რაც თავისთავად უარყოფითად აისახება ქვეყანაში ბიზნესის განვითარებაზე. აღსანიშნავია, რომ 2010 წლის განმავლობაში გადამხდელზე დაკისრებული გადასახადისა და სანქციების შემცირების მიზნით 250-მდე საგადასახადო შეთანხმება გაფორმდა.

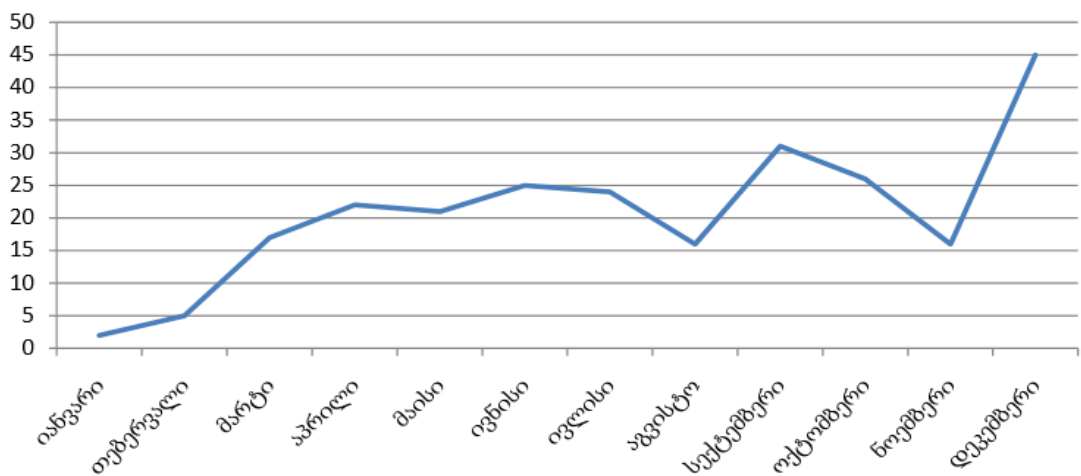

წყარო: შემოსავლების სამსახური 2010 წელს გაწეული საქმიანობის ანგარიში მნიშვნელოვნად გამარტივდა საზღვარზე განხორციელებული პროცედურებიც. როგორც ცნობილია, საქართველო აქტიურ საქმიანობას ეწევა ექსპორტ-იმპორტის მიმართულებით. ამ მხრივ მეტად რთული და ხანგრძლივი პროცედურები ხორციელდებოდა რეფორმის განხორციელებამდე. რეფორმის შემდგომ უფრო გამარტივდა საქონლის საბაჟო საზღვარზე კვეთის ოპერაციები, უფრო ოპერატიულად და რაც მთავარია ნაკლები დანახარჯებით ხორციელდება სასაქონლო ოპერაციები. გარდა ამისა, „საქართველო 2010 წლიდან შეუერთდა საერთაშორისო კონვენციას საგადასახადო საკითხებში ორმხრივი ადმინისტრაციული დახმარების თაობაზე, რაც საშუალებას იძლევა გაუმჯობესდეს საერთაშორისო კომპანიების საგადასახადო ხარისხი, რიგ ქვეყნებთან (აშშ, აზერბაიჯანი, ბელგია, დანია, ფინეთი, შვედეთი, საფრანგეთი, ისლანდია, იტალია, ნიდერლანდები, ნორვეგია, პოლონეთი, შვედეთი, დიდი ბრიტანეთი, უკრაინა) საგადასახადო საკითხებზე ინფორმაციის ურთიერთგაცვლის მიზნით" (შემოსავლების სამსახური, 2010).

შემოსავლების სამსახურმა ასევე დანერგა, საქონლის წინასწარი დეკლარირების სისტემა, რომელიც ხელს უწყობს საქონლის შემოტანამდე ოფიციალური კონტროლის განხორციელებას, ხოლო შემოტანის მომენტში ხორციელდება მარტივი ხანმოკლე პროცედურები. აღნიშნული რეფორმა, მნიშვნელოვნად ამცირებს სასაზღვრო-გამშვებ პუნქტებზე პროცედურების ხანგრძლივობასა და დანახარჯებს. გარდა ამისა, ჩამოყალიბდა ექსპერტიზის დეპარტამენტი და კინოლოგიური სამსახური, გაიხსნა არა ერთი გაფორმების ეკონომიკური ზონა და ა. შ ამ და სხვა ღონისძიებების განხორციელებით შემოსავლების სამსახური ხელს უწყობს რომ უფრო მოქნილი, ეფექტური და ოპერატიული გახდეს გადასახადის გადამხდელსა და საგადასახადო ორგანოებს შორის ურთიერთობა, რომელიც ერთის მხრივ გადამხდელებს დაეხმარება ზუსტად განსაზღვრონ მათი საგადასახადო ვალდებულება და მეორე მხრივ საგადასახადო ორგანოებს გაუადვილებს საგადასახადო ადმინისტრირების ეფექტურად განხორციელებას და საჭირო ოდენობის რესურსების ბიუჯეტში მობილიზებას, მცირე დროში და ნაკლები დანახარჯებით.

„საქართველოს მომავალი დგას ჩვენი ქვეყნის მეწარმეების, მცირე და საშუალო ბიზნესმენების, საქმიანი ადამიანების მხრებზე. ჩვენი ხალხისა და ქვეყნისათვის ყველაზე დიდი მნიშვნელობა აქვს, რომ მეწარმე თავს გრძნობდეს კომფორტულად, გრძნობდეს, რომ მის ირგვლივ არის შექმნილი სამართლიანი ეკონომიკური გარემო და იყოს თავდაჯერებული. მხოლოდ ასეთი მეწარმე შეძლებს ქვეყნის ეკონომიკის განვითარებას, ახალი სამუშაო ადგილების შექმნას, მეტი ეროვნული დოვლათის დაგროვებას" (შემოსავლების სამსახური, 2010). ხელისუფლება ძირითადად აქცენტს აკეთებს ბიზნეს სუბიექტებისათვის სტაბილური ბიზნეს გარემოს შექმნაზე, სწორედ ამიტომ ბევრი მნიშვნელოვანი რეფორმა გატარდა მიკრო და მცირე ბიზნესის გაუმჯობესებისა და განვითარებისკენ. 2010 წელს გატარებულმა საგადასახადო ცვლილებებმა მკაფიოდ განსაზღვრა მიკრო და მცირე ბიზნესის დეფინიციები და დაადგინა შეღავათიანი და მარტივად ადმინისტრირებადი გადასახადის განაკვეთები. მიკრო ბიზნესი გათავისუფლდა საშემოსავლო გადასახადის გადახდის ვალდებულებისგან. რაც შეეხება, მცირე ბიზნესის მქონე მეწარმეებს, 2018 წლის 1-ლი ივლისიდან ძირეული ცვლილებები გატარდა ამ მიმართულებით. მცირე ბიზნესის წახალისებისა და ხელშეწყობის მიზნით ძალაში შევიდა დაბეგვრის შედავათიანი რეჟიმი, რომელიც 120 000-მდე მეწარმე-ფიზიკურ პირს შეეხო. ახალი საგადასახადო რეჟიმი ერთის მხრივ, აწესებს მინიმალურ საგადასახდო ტვირთს მცირე ბიზნესისათვის, რომელიც განისაზღვრა 1 %-ით ნაცვლად 5 %-სა, ხოლო მეორეს მხრივ, ზრდის ერთობლივი შემოსავლის ზედა ზღვარს, კერძოდ ახალი რეფორმის შესაბამისად მცირე ბიზნესის სტატუსი მიენიჭებათ მეწარმე ფიზიკურ პირებს, რომელთა ერთობლივი შემოსავალი კალენდარული წლის განმავლობაში იქნება 500 000 ლარამდე, ნაცვლად მანამდე არსებული 100 000 ლარისა. გარდა ამისა 500 000 ლარზე გადაჭარბების შემთხვევაში მეწარმე ფიზიკურ პირს მცირე ბიზნესის სტატუსი უნარჩუნდება 2 კალენდარული წლის განმავლობაში. აღნიშნული რეფორმა ასევე უზრუნველყოფს მეწარმეებისათვის გამარტივებული საგადასახადო აღრიცხვის წარმოებას, რაც მინიმუმამდე შეამცირებს საგადასახადო ადმინისტრირების ხარჯებს. გარდა ამისა, მცირე ბიზნესის სტატუსის გაუქმების საფუძველი აღარ იქნება მეწარმე ფიზიკური პირის დამატებული ღირებულების გადასახადის გადამხდელად რეგისტრაცია. ამასთან, მცირე ბიზნესის სტატუსის მქონე პირს აღარ აქვს მიმდინარე გადასახდელების გადახდის ვალდებულება და დეკლარირება ხორციელდება ყოველთვიური მარტივი დეკლარაციის ფორმის მეშვეობით.

აღნიშნული რეფორმის შედეგად მნიშვნელოვნად გაიზრდება მთლიან შიდა პროდუქტში მცირე ბიზნესის წილი. საგადასახადო დაბეგვრის შედავათიანი რეჟიმი მნიშვნელოვანი სტიმულია მცირე ბიზნესის შემდგომი განვითარებისთვის. აღნიშნულ რეფორმას დადებითი ეფექტი აქვს საგადასახადო შემოსავლებისთვისაც, იმ თვალსაზრისით, რომ აქამდე არსებული ბიზნესები რომლებიც ე.წ "იატაკქვეშა ეკონომიკას" ეწეოდნენ, უკვე კანონიერად დაიწყებენ სამეწარმეო საქმიანობას, რათა მათი საგადასახადო ვალდებულება საგრძნობლადაა შემცირებული და იგი აღარ წარმოადგენს დამატებით ტვირთს მათი ბიზნესისათვის. გარდა ამისა, რაც უფრო მაღალი ბრუნვის საწარმოები ისარგებლებენ ამ სტატუსით მით უფრო ექნებათ მათ სტიმული რომ უფრო გააფართოონ და განავითარონ მათი ბიზნესი, რაც შექმნის დამატებით სამუშაო ადგილებს და ხელს შეუწყობს მთლიანი შიდა პროდუქტის ზრდას.

საგადასახადო სტიმულირების ღონისძიებების ძირითადი მიზანია სტაბილური ბიზნეს გარემოს შექმნა მეწარმე სუბიექტებისათვის, რათა მათ ეფექტიანად შეძლონ სამეწარმეო საქმიანობის განხორციელება და შესაბამისად ხელი შეუწყონ სახელმწიფო ბიუჯეტში საჭირო ოდენობის ფინანსური რესურსების მობილიზებას. ამ მხრივ ძირეული ცვლილებები შევიდა საგადასახადო კოდექსში მოგების გადასახადის მიმართულებით. ჯერ კიდევ 2015 წელს დაანონსდა მოგების ე.წ. „ესტონური" მოდელით დაბეგვრის რეჟიმის შემოღება, რომელიც ძალაში 2017 წლის 1-ლი იანვრიდან შევიდა. აღნიშნულ მოდელთან დაკავშირებით, მის ამოქმედებამდეც და შემდგომაც მწვავე დებატები მიმდინარეობდა; მთავარი განხილვის საგანი იყო ის თუ აღნიშნული რეფორმა რა გავლენას მოახდენდა მეწარმე სუბიექტების საქმიანობაზე და ამასთანავე ბიუჯეტის საგადასახადო შემოსავლებზე. გამომდინარე იქიდან, რომ მოგების გადასახადი ნებისმიერი ბიზნეს სუბიექტისთვის წარმოადგენს მაგისტრალურ გადასახადს, მნიშვნელოვანია იმ თავისებურებების გამოყოფა, რაც განასხვავებს „ესტონურ" მოდელს აქამდე არსებული მოდელისაგან. პრინციპულ სიახლეს ახალ მოდელთან დაკავშირებით წარმოადგენს ის გარემოება, რომ მეწარმე სუბიექტს მოგების გადასახადის გადახდის ვალდებულება წარმოექმნება მხოლოდ მოგების განაწილების მომენტში, რაც მათ საშუალებას აძლევთ დაზოგონ თანხები მოგების მიღებიდან მის განაწილებამდე და მიმართონ აღნიშნული სახსრები რეინვესტირებაზე, ანუ მოახდინონ ფინანსური რესურსების კაპიტალად ტრანსფორმაცია. მოგების გადასახადით დაბეგვრის „ესტონური" მოდელის, მიხედვით დაბეგვრის ობიექტს წარმოადგენს:

➢ განაწილებული მოგება ( მათ შორის ფასთაშორისი სხვაობები);

➢ არაეკონომიკური ხარჯები და გადახდები;

➢ უსასყიდლო მიწოდება (მ.შ. საქონლის მიწოდება, მომსახურების გაწევა და ფულადი სახსრების გადაცემა);

➢ წარმომადგენლობითი ხარჯი ზღვრული ოდენობის ზევით.

„ესტონური" მოდელის დადებით მხარეებზე თუ ვისაუბრებთ, იგი ხელს უწყობს ქვეყანაში რეინვესტირების პროცესს და ბიზნეს სუბიექტებს საშუალებას აძლევს დაზოგონ ფინანსური რესურსები; ეს უკანასკნელი მეტად მნიშვნელოვანია დამწყები ბიზნესისთვის, რათა სამეწარმეო საქმიანობის გააქტიურებამდე მათ შეძლონ საჭირო ოდენობის ფინანსური რესურსების მობილიზება.

როგორც ზევით, აღვნიშნეთ თითოეული საკანონდებლო ცვლილების მიზანი უნდა იყოს ქვეყანაში სამეწარმეო საქმიანობის ხელშეწყობა და ეკონომიკური ზრდის სტიმულირება, თუმცა ყურადღების მიღმა არ უნდა დარჩეს საგადასახადო ვალდებულების შესრულების დონე, თითოეულმა გატარებულმა რეფორმამ თუ სტიმულირების ღონისძიებამ ხელი უნდა შეუწყოს სახელმწიფო ბიუჯეტში საჭირო ოდენობის ფინანსური რესურსების მობილიზებას, რაც აუცილებელია ნებისმიერი სახელმწიფოსთვის მასზე დაკისრებული ვალდებულებების შესასრულებლად. „ესტონური" მოდელის გავლენა, ყველა ცალსახად აღნიშნავს, რომ ბიუჯეტის საგადასახადო შემოსავლებისთვის უარყოფითია, რადგან რეფორმის ამოქმედებიდან საბიუჯეტო შემოსავლები წინა წელთან შედარებით 650 მლნ ლარამდე შემცირდა. დიაგრამა 3-ზე თვალსაჩინოდაა წარმოდგენილი რეფორმის ამოქმედებით გამოწვეული შედეგები.

**დიაგრამა 3.** მოგების გადასახადის დინამიკა 2016-2019 წლების სახელმწიფო ბიუჯეტის შემოსავლებში (ათასი ლარი)

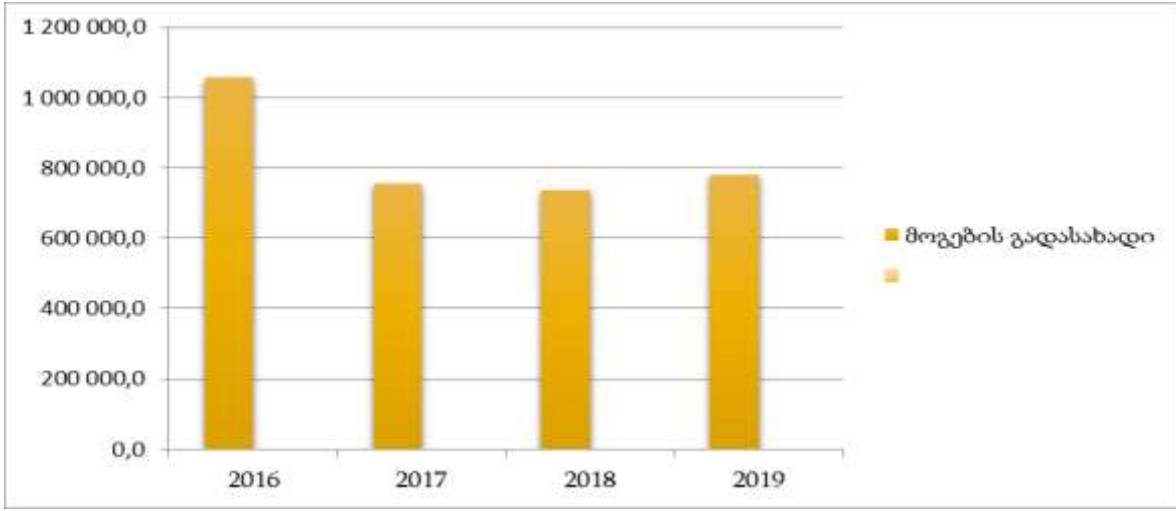

**წყარო:** შედგენილია ავტორის მიერ საქართველოს ფინანსთა სამინისტროს მონაცემებზე დაყრდნობით ამ დეფიციტის დასაფინანსებლად თავიდან ბიუროკრატიული ხარჯების შემცირება იგეგმებოდა მაგრამ საბოლოოდ ეს ტვირთი აქციზის გადასახადის გადამხდელებზე გადავიდა და შესაბამისად 380 მილიონითაა გაზრდილი 2017 წელს აქციზის გადასახადით მობილიზებული სახსრები წინა წელთან შედარებით. ანუ ის რაც მოგების გადასახადით „დაუტოვა" სახელმწიფომ კერძო სექტორს აქციზის გადასახადით „ამოიღო". უარყოფითი ფისკალური ეფექტის მიუხედავად, მოგების „ესტონური მოდელით" დაბეგვრის რეჟიმი უდიდესი მიღწევაა ჩვენი ქვეყნის ეკონომიკისათვის, იგი გაზრდის ეკონომიკაში შიდა ინვესტიციების მოცულობას და შესაბამისად წახალისებს ეკონომიკურ აქტივობას გრძელვადიან პერიოდში. აქვე დავამატებ, რომ აღნიშნული რეფორმა არ ეხება კომერციულ ბანკებს, მიკროსაფინანსო ორგანიზაციებს, საკრედიტო კავშირებს, სადაზღვევო კომპანიებს, ლომბარდებს (საგადასახადო კოდექსის გარდამავალი დებულებებით განსაზღვრულ ვადამდე), რაც ვფიქრობ არა კონკურენტულ გარემოს ქმნის კერძო სექტორში. ყველაფრის მიუხედავად, საგადასახადო კანონმდებლობაში აღნიშნული რეფორმა საშუალოვადიან და გრძელვადიან პერიოდში სამეწარმეო საქმიანობის გააქტიურებასა და ეკონომიკურ ზრდას გვიწინასწარმეტყველებს.

სახელმწიფო ბიუჯეტის საგადასახადო შემოსავლების სტატისტიკას თუ გადავხედავთ ყველაზე მაღალი წილით დამატებული ღირებულების გადასახადი გამოირჩევა, ამიტომაც ამ მიმართულებითაც აღმასრულებელი ხელისუფლების მხრიდან აქტიური ღონისძიებები ხორციელდება. უახლოესი რეფორმა რაც დამატებული ღირებულების გადასახადს შეეხო გულისხმობს გადასახადის გადამხდელის მიერ ზედმეტად გადახდილი გადასახადის ავტომატურ რეჟიმში უკან დაბრუნებას. რეფორმა 2019 წლის იანვრიდან ამოქმედდა და ბიზნესს არა მარტო დროის არამედ ადამიანური რესურსის დაზოგვის შესაძლებლობას მისცემს. რაც უფრო დროულად მიიღებს ბიზნესი თავის კუთვნილ თანხას მით უფრო უწყვეტად იფუნქციონირებს სამეწარმეო საქმიანობაც და უფრო ლიკვიდური იქნება ბიზნესი და შესაბამისად ნაკლებად იქნება იგი სესხზე დამოკიდებული და საკუთარი სახსრებით შეძლებს საქმიანობას. აღნიშნული რეფორმა მართლაც, ერთ-ერთი წინ გადადგმული ნაბიჯია საგადასახადო სისტემაში, იგი მსოფლიოს ბევრი განვითარებული ქვეყნის პრაქტიკას ეყრდნობა. თანამედროვე ეტაპზე დროითი, ფინანსური და ადამიანური რესურსები, ეს ის სამი მნიშვნელოვანი საყრდენია რომელმაც უნდა განსაზღვროს ბიზნესის მომავალი, ამიტომაც რაც უფრო ოპერატიულად, ნაკლებ დროში, მინიმალური დანახარჯებით და ძალისხმევით წარიმართება სამეწარმეო საქმიანობა მით უფრო ძლიერი იქნება ქვეყნის ეკონომიკაც.

აღმასრულებელი ხელისუფლების მთელი ძალისხმევა მიმართული უნდა იყოს საგადასახადო ვალდებულების ნებაყოფლობით შესრულების, „ჩრდილოვანი" ეკონომიკის აღმოხვერის, სამართალდარღვევათა პრევენციის და რაც მთავარია, ბიზნესისთვის ხელსაყრელი გარემოს შექმნის და მისი გაუმჯობესებისაკენ. საგადასახადო პოლიტიკა და ადმინისტრირება ჰარმონიულად უნდა მუშაობდეს; საგადასახადო სისტემა ქვეყნის ეკონომიკური სისტემის შემადგენელი ნაწილია და მას სტრატეგიული მნიშვნელობა ენიჭება, როგორც ქვეყნის ეკონომიკური ასევე ფისკალური პოლიტიკით განსაზღვრული ამოცანების განხორციელებაში.

სახელმწიფოს მიერ გატარებული თითოეული ღონისძიება ხელს უნდა უწყობდეს საბიუჯეტო შემოსავლების მობილიზებას საზოგადოებრივი კეთილდღეობის ხელშეწყობით, უნდა შექმნას ხელსაყრელი გარემო ნებაყოფლობითი კანონ შესაბამისობის წასახალისებლად. „აუცილებელია გასიგრძეგანებულ იქნეს, რომ ყოველი გადაუხდელი ლარი არის არა მხოლოდ საგადასახადო კანონდარღვევა არამედ ეს არის საქართველოს თითოეული მოქალაქისთვის, მისი მომავალი თაობისთვის მოპარული, დაკარგული რესურსი" (შემოსავლების სამსახური, 2016). საგადასახადო სისტემის ეფექტურად მუშაობა გადასახადის გადამხდელთა საგადამხდელო კულტურის ამაღლებაზე, კეთილდღეობასა და განვითარებაზეა დამოკიდებული, ჩვენ ყველას ერთად, სწორი და თანმიმდევრული ნაბიჯებით, ღირსეულად შეგვიძლია ქვეყნის წინაშე ჩვენზე დაკისრებული ძალიან მნიშვნელოვანი და საპასუხისმგებლო ფუნქციების წარმატებით შესრულება.

## 2.2 საგადასახადო შემოსავლების ფორმირების უცხოური გამოცდილება და მისი დანერგვის პერსპექტივები საქართველოში

ქვეყნების ბიზნეს გარემოს შეფასებისას გადასახადების დონის განსაზღვრას მნიშვნელოვანი ადგილი უჭირავს. ქვეყნების შედარება გადასახადების დონეების მიხედვით საკმაოდ რთულია, რადგან არ არსებობს რაიმე აღიარებული საერთაშორისო რეიტინგი, რომლის მეთოდოლოგია ყველასთვის მისაღები იქნება.

ამა თუ იმ საგადასახადო რეფორმის გატარების დროს გათვალისწინებული უნდა იქნას სხვადასხვა ქვეყნებში არსებული პრაქტიკა. თანამედროვე ეტაპზე სახელმწიფოები იზიარებენ და ითვალისწინებენ სხვა ქვეყნების გამოცდილებას და ეროვნული თავისებურებების შესაბამისად იყენებენ მას. გამომდინარე იქედან, რომ საქართველო ღია ეკონომიკის პატარა ქვეყნების რიგს მიეკუთვნება, ხშირად ხდება სხვადასხვა ქვეყნებში არსებული საგადასახადო სისტემების გარკვეული ელემენტების ჩვენი ქვეყნის საგადასახდო სისტემაში დანერგვა. უცხოური გამოცდილების გაზიარება საქართველოსთვის მეტად მნიშვნელოვანია, აქამდე არსებული საგადასახადო დაბეგვრის მოდელები, რომლებიც უცხოურ გამოცდილებას ეფუძნება დიდი წარმატებით ფუნქციონირებს.

ჩვენი ქვეყანა კვლავაც განაგრძობს განვითარებულ ქვეყნებში არსებული საგადასახადო მოდელების დანერგვას, მაგრამ ყოველივე ამას ღრმა ანალიზი, კვლევა ძიება ესაჭიროება რადგან ამა თუ იმ ქვეყანაში წარმატებით მოქმედი საგადასახადო სისტემა შესაძლოა ისევე კარგად არ მოერგოს ჩვენს ქვეყანას როგორც სხვა ქვეყნებში. განვითარებული ქვეყნების უმეტესობაში, როგორიც არის გერმანია, შვეიცარია, საფრანგეთი, პროგრესული საგადასახადო სისტემები მოქმედებს, რაც იმას ნიშნავს, რომ რაც მეტია დასაბეგრი ბაზა მით მაღალია საგადასახადო განაკვეთი და შესაბამისად ყველას თანაბარ პირობებში უწევს საგადასახადო ვალდებულების შესრულება.

საქართველოში ამჯამად უფრო მეტად დაბეგვრის პროპორციული სისტემა გამოიყენება, რაც იმას ნიშნავს რომ საგადასახადო ბაზის მიუხედავად ყველასთვის ერთიანი საგადასახადო განაკვეთია დაწესებული; ასეთი სისტემის დროს სტიმული არ ქრება, რომ ადამიანმა იმუშაოს და მიიღოს მეტი შემოსავალი, ვინაიდან, მისი დაბეგვრის რეჟიმი, პროცენტული მაჩვენებელი არ იცვლება, შესაბამისად, მიიღებს მეტ შემოსავალს და იგივე განაკვეთით დაიბეგრება, მაგრამ დაბეგვრის ამგვარი სისტემა დისკრიმინაციულ დამოკიდებულებას ქმნის გადასახადის გადამხდელებში. გამომდინარე იქედან, რომ მაგალითად: 200 და 2000 ლარიანი შემოსავლის მქონე პირები ერთნაირი 20%-იანი განაკვეთით იბეგრებიან გარკვეულწილად უსამართლობის განცდა ჩნდება გადასახადის გადამხდელებს შორის. იმ ქვეყნებში სადაც პროგრესული დაბეგვრა მოქმედებს ძირითადად გამოიყენება პროცენტული შკალა და საპროცენტო განაკვეთები შემოსავლების მიხედვით ნაწილდება თითოეულ ინტერვალზე, ასე მაგალითად:

- 0-600 ლარამდე  - 0%
- 600-დან 100 ლარამდე - 10%
- 1000-დან 1800 ლარამდე - 20%
- 1800-დან 3000 ლარამდე - 35%
- 3000 ლარის ზევით - 45%

"იმისთვის რომ დაქირავებული მუშების, მეწარმეების და სახელმწიფო ბიუჯეტის შემოსავლები მაქსიმალური ეფექტურობით იქნას გამოყენებული, საჭიროა საგადასახადო სისტემის რეორგანიზაცია, როგორც ეს უკვე მოხდა განვითარებულ სახელმწიფოებში. საკმარისი არაა მხოლოდ საშემოსავლოს გადასახადის პროგრესული შკალის შემოდება, ასევე საჭიროა პროგრესული გადასახადის დაწესება მოგებაზე, უძრავ ქონებაზე, ფუფუნების ძვირადღირებულ ნივთებზე და ა.შ" (პროგრესული დაბეგვრა - სამართლიანობისთვის, განვითარებისთვის და სტაბილურობისთვის, 2017).

აღსანიშნავია ისიც, რომ დაბეგვრის პროგრესული სისტემის პირობებში ნებისმიერ ადამიანს უქრება სტიმული მაღალი შემოსავლის მიღების ან სხვაგვარად, რომ ვთქვათ ემჯეს „ხვრელს" ნაკლები გადასახადის გადახდისთვის. წარსულში საქართველოშიც იყო საუბარი დაბეგვრის პროგრესული სისტემის დანერგვაზე, მაგრამ ჩვენს ქვეყანაში არსებულ ბიზნეს გარემოს პროპორციული საგადასახადო სისტემა უფრო შეესაბამება, რადგან პროგრესული საგადასახო სისტემა გაზრდის შემოსავლების დამალვის და ე.წ. „ჩრდილოვანი ეკონომიის" ჩამოყალიბებას.

აღსანიშნავია ისიც რომ ზოგიერთ განვითარებულ ქვეყანაში არსებობს ე.წ. „ფუფუნების გადასახადი". ისტორიულად ფუფუნებაზე გადასახადი წარმოადგენს შუალედურ როლს, რომელსაც შეუძლია მიიყვანოს სისტემა თანამედროვე, განვითარებულ პროგრესულ გადასახადამდე, რომელიც მიღებულია სხვადასხვა ქვეყნებში გარკვეული ფორმით. თავის დროზე ასეთი გადასახადის შემოდება საზოგადოებისთვის იყო ახალი პროგრესული საშემოსავლო გადასახადის შემოდებისთვის მზადება. აღნიშნულმა გადასახადმა შეიძლება საგადასახადო სისტემის სასარგებლო კორექტირებას შეუწყოს ხელი იმ ქვეყნებში სადაც საგადასახადო წნეხი მთლიანად საშუალო და დაბალ ფენას აწვება. ბრიტანეთში, მაგალითად, სადაც ყველაზე ადრე დაიწყეს ამ თემაზე საუბარი, კანონის ადმინისტრაცია ევალება არა ცენტრს, არამედ ადგილობრივ მუნიციპალიტეტებს, რაც აადვილებს გადასახადების ამოღებას. ფუფუნების გადასახადი არსებობს ასევე მექსიკაში, ჩილეში, ავსტრალიაში, ალჟირში, უნგრეთში და ბევრ სხვა ქვეყანაში. რატომ არ შეიძლება მსგავსი გადასახადის შემოდება საქართველოში? ამ გადასახადის შემოდებას პოლიტიკური ნება ჭირდება, რაც განპირობებული იქნება ქვეყანაში კატასტროფული სოციალურ-ეკონომიკური უთანასწორობის შემცირების მოტივით, ასევე იმ გარემოებით რომ საგადასახადო ტვირთი ძირითად საშუალო და ღარიბ ფენას არ დააწვეს და ფინანსური ელიტები დაიბეგრონ საკუთარი ქონების პროპორციულად.

აღსანიშნავია ისიც, რომ საფრანგეთსა და შვეიცარიაში წარმატებით ფუნქციონირებს დიფერენცირებული საგადასახადო განაკვეთები დასაბეგრი ობიექტების მიხედვით. კერძოდ, შვეიცარიაში დამატებული ღირებულების გადასახადი პირველადი მოთხოვნილების საქონელზე: საკვებზე, ტანსაცმელზე, სამედიცინო საქონელზე, გაზეთებზე, წიგნებზე (2,4%), საცხოვრებელ მომსახურებაზე (3,6%), სტანდარტულ განაკვეთთან (7,6%) შედარებით საგრძნობლად დაბალია (სინგაპური, ესტონეთი, შვეიცარია: სამიზნე მოდელები ქართული ტრანსფორმაციისთვის, 2012).
საფრანგეთში სტანდარტული განაკვეთი დამატებული ღირებულების გადასახადზე 20%-ია, ხოლო კვების პროდუქტებზე და სხვა პირველადი მოხმარების საქონელზე 5,5%-ია, სამკურნალო პროდუქციასა და მედიკამენტებზე კი 2,1%. დიდი ბრიტანეთის მაგალითს თუ ავიღებთ სურსათი, წყალი, მედიკამენტები, საზოგადოებრივი ტრანსპორტი მთლიანად გათავისუფლებულია დამატებული ღირებულების გადასახადისგან.

საგადასახდო განაკვეთების დიფერენციაცია, მით უფრო დამატებული ღირებულების გადასახადისა, რომელიც ერთ-ერთ მაგისტრალურ გადასახადს წარმოადგენს ვფიქრობთ, მნიშვნელოვანი წინ გადადგმული ნაბიჯი იქნება ჩვენი ქვეყნის საგადასახადო სისტემაში. გამომდინარე იქედან, რომ დამატებული ღირებულების გადასახადის უშუალო გადამხდელი საქონლისა და მომსახურების საბოლოო მომხმარებელია იგი ხელს შეუწყობს, რომ საზოგადოების დაბალშემოსავლიანმა ნაწილმა ხელმისაწვდომ ფასად მიიღონ პირველადი მოხმარების საქონელი და მომსახურება.

ნებისმიერ სახელმწიფოში საგადასახადო სისტემა იმგვარად უნდა იყოს ჩამოყალიბებული, რომ იგი როგორც ხალხისთვის ასევე ეკონომიკისთვისაც იყოს სასარგებლო. საქართველოში საგადასახადო ცნობიერება ერთგვარი გაგრძელებაა იმ საბჭოთა ცნობიერების, რომელიც მოქალაქეებს არა ღია და კანონიერი, არამედ ფარული გზით ყველაფრის მიღწევის გარანტიას აძლევდა. ამიტომაც ამა თუ იმ საგადასახადო რეფორმის გატარებისას აუცილებლად უნდა იქნას გათვალისწინებული ქვეყანაში არსებული საგადასახადო კულტურა და ზოგადად საზოგადოების დამოკიდებულება სახელმწიფო ვალდებულებების შესრულების დროს.

## დასკვნები და წინადადებები:

- საგადასახადო შემოსავლების სრულყოფის საკითხი მეტად აქტუალურია თანამედროვე ეტაპზე; ყველა სახელმწიფო ცდილობს, რომ მაქსიმალურად ეფექტიანად წარმართოს საგადასახადო პოლიტიკა ქვეყანაში, რადგან დიდ

- წილად მასზეა დამოკიდებული ქვეყანაში მიმდინარე ეკონომიკური პროცესების სწორად წარმართვა. უნდა მოხდეს საგადასახადო სისტემის კომპლექსური ანალიზი და გადაიდგას ქმედითი ნაბიჯები, რომელიც დადებითად აისახება სახელმწიფო ბიუჯეტის შემოსავლებზე.
- საგადასახადო სისტემის სრულყოფისათვის ასევე აუცილებელია დაიხვეწოს საგადასახადო კანონმდებლობა, გადასახადის გადამხდელებისთვის უფრო გასაგებად და არაორაზროვნად უნდა ჩამოყალიბდეს საგადასახადო კოდექსი, რადგან ზუსტად იქნეს განსაზღვრული საგადასახადო ორგანოებისა და გადასახადის გადამხდელების უფლება-მოვალეობები და რაც ყველაზე მთავარია, გადასახადის გადამხდელთა საგადასახადო ვალდებულება.
- სახელმწიფო სამეწარმეო სუბიექტების საქმიანობის წასახალისებლად ხშირად ატარებს მასტიმულირებელ ღონისძიებებს, შეღავათებს უწესებს გადასახადის გადამხდელებს, აწესებს დაბალ განაკვეთებს მცირე ბიზნესებისთვის და ა.შ. თუმცა, თითოეულ გატარებულ რეფორმას სჭირდება ღრმა ანალიზი და კვლევა, თუ რამდენად ეფექტიანად იმუშავებს აღნიშნული რეფორმა ჩვენს ქვეყანაში, მნიშვნელოვანია ასევე განისაზღვროს საგადასახადო კულტურა რა დონეზე ქვეყანაში განვითარებული და რამდენად შეუწყობს ხელს გატარებული რეფორმები საბიუჯეტო შემოსავლების ფორმირებას.
- საგადასახადო სისტემის ეფექტიანად წარმართვისათვის ასევე აუცილებელია განხილული იქნას სხვა ქვეყნებში არსებული პრაქტიკა და მისი დანერგვის პერსპექტივები, როგორიცაა, პროგრესული დაბეგვრის სისტემა, რაც გულისხმობს დასაბეგრი ბაზის გაზრდის პროპორციულად საგადასახადო განაკვეთის ზრდას, რაც უფრო სამართლიანია და ხელს უწყობს უფრო მეტი ფინანსური რესურსების მოზიდვას ბიუჯეტში, მაგრამ ხაზგასასმელია ის ფაქტიც, რომ დაბეგვრის პროგრესული სისტემა ხელს შეუწყობს „ჩრდილოვანი" ეკონომიკის ჩამოყალიბებას და გაზრდის გადასახადებისგან თავის არიდების და დამალვის ფაქტებს.
- გარდა ამისა, კიდევ ერთხელ გავუსვამთ ხაზს გადასახადების განაკვეთების დიფერენციაციის საკითხს. ვფიქრობ, ჩვენი ქვეყნის ეკონომიკისთვის მნიშვნელოვანი იქნება რეფორმის გატარება ამ მიმართულებით, რადგან იგი ხელს შეუწყობს სოციალური პრობლემების მოგვარებას ქვეყანაში და მოსახლეობის დაბალ შემოსავლიანი ნაწილისთვის სასარგებლო იქნება.

გამოყენებული ლიტერატურა:

mthavrobis-administratsiuli-kharjis-shemtsirebam-da-aqtsizis-ganakvethis-zrdam-daabalansa;

29. მცირე ბიზნესის სტატუსის 500 000 ლარამდე გაზრდის მიუხედავად, დღგ-ს ლიმიტი კვლავ 100 000 ლარია, 2018; http://www.tabula.ge/ge/story/129815-mcire-biznesis-statusis-500000-laramde-gazrdis-miuxedavad-dghg-s-limiti-kvlav-100 000;

30. საგადასახადო ცვლილებები მეწარმეებს მცირე ბიზნესის სტატუსი 500 000 ლარის შემოსავლის შემთხვევაში მიენიჭებათ, 2018; https://1tv.ge/news/sagadasakhado-cvlilebit-mewarmeebs-mcire-biznesis-statusi-wliuri-500-000-laris-shemtkhvevashi-mienichebat/;

31. დღგ-ს უკან დაბრუნება, 2019; https://www.bm.ge/ka/video/dgg-s-ukan-dabruneba/11882;

32. საქართველოს ფინანსთა სამინისტრო, https://mof.ge/ ;

33. შემოსავლების სამსახური, https://www.rs.ge//

34. რომელ ქვეყანაშია ყველაზე დაბალი გადასახადები, 2017; https://forbes.ge/news/3234/romel-qveynebSia-dabali-gadasaxadebi ;

35. სხვადასხვა ქვეყნის საგადასახადო სისტემები, 2019. http://www.worldwide-tax.com/history_of_tax.asp

36. შემოსავლების სამსახური, 2010 წელს გაწეული საქმიანობის ანგარიში, 2011; https://www.rs.ge/common/get_doc.aspx?id=7129;

37. შემოსავლების სამსახური, 2016 წლის ანგარიში, https://www.rs.ge/common/get_doc.aspx?id=10180

38. ბოლქვაძე, ბ. (2012) „ფინანსური მენეჯმენტი", ბათუმი;

39. საქართველოს საგადასახადო კოდექსი, 2019, https://matsne.gov.ge/ka/document/view/1043717
31